\documentclass[a4paper, 12pt]{article}

\usepackage[top=2cm, bottom=2cm, left=2.5cm, right=2.5cm]{geometry}
\usepackage{amssymb}
\usepackage{amsmath}
\usepackage{amsfonts}
\usepackage{bm}
\usepackage{hyperref}
\usepackage[natbibapa, nosectionbib]{apacite}
\usepackage{natbib}
\usepackage{setspace}
\usepackage{xcolor}
\usepackage{changes}
\usepackage{multirow}
\usepackage{longtable}
\usepackage{supertabular} 
\usepackage{dcolumn}
\usepackage{pdflscape}
\usepackage[shortlabels]{enumitem}
\usepackage{tabularx,calc}
\usepackage{booktabs}
\usepackage{acro}
\usepackage{setspace}
\usepackage{afterpage}
\usepackage{lscape}
\usepackage{mathtools}
\usepackage{pst-plot, pstricks} 
\usepackage{fancybox,amssymb,color}
\usepackage{subfigure}
\usepackage{framed}
\usepackage{scalerel}[2016-12-29]
\usepackage{titling}
\usepackage[bottom]{footmisc} 
\usepackage[]{authblk}
\usepackage[title]{appendix}
\usepackage[position=bottom, justification=centering, font=footnotesize,labelfont=bf]{caption}
\usepackage{makecell}
\usepackage{float}

\newcommand{\spanS}{\text{span}}
\newcommand*\diff{\mathop{}\!\mathrm{d}}

\newtheorem{preremark}{Remark}
\newenvironment{remark}%
  {\begin{preremark}\upshape}{\end{preremark}}
\DeclareMathOperator*{\argmin}{arg\,min}

\makeatletter
\newcommand{\subalign}[1]{%
	\vcenter{%
		\Let@ \restore@math@cr \default@tag
		\baselineskip\fontdimen10 \scriptfont\tw@
		\advance\baselineskip\fontdimen12 \scriptfont\tw@
		\lineskip\thr@@\fontdimen8 \scriptfont\thr@@
		\lineskiplimit\lineskip
		\ialign{\hfil$\m@th\scriptstyle##$&$\m@th\scriptstyle{}##$\crcr
			#1\crcr
		}%
	}
}
\makeatother

\definecolor{Diceblue}{RGB}{52,116,181}

\hypersetup{
	colorlinks = true,
	linkcolor=Diceblue,
	citecolor=Diceblue,
	urlcolor=Diceblue
}

\newcolumntype{B}{@{\extracolsep{-0.0pt}}c@{\extracolsep{0.0pt}}}
\newcolumntype{z}{@{\extracolsep{0.3cm}}c@{\extracolsep{0.3cm}}}

\begin{document}

	
\setlength{\abovedisplayskip}{3pt}
\setlength{\belowdisplayskip}{12pt}
	
	
\setlength{\jot}{0.5cm}

\begin{titlingpage}

\title{A Sparse Grid Approach for the Nonparametric Estimation of High-Dimensional Random Coefficient Models}

\author{Maximilian Osterhaus$^\dagger$}
\affil{University of Groningen}

\date{February 2022}
	

\maketitle

\begin{abstract}
A severe limitation of many nonparametric estimators for random coefficient models is the exponential increase of the number of parameters in the number of random coefficients included into the model. This property, known as the curse of dimensionality, restricts the application of such estimators to models with moderately few random coefficients. This paper proposes a scalable nonparametric estimator for high-dimensional random coefficient models. The estimator uses a truncated tensor product of one-dimensional hierarchical basis functions to approximate the underlying random coefficients' distribution. Due to the truncation, the number of parameters increases at a much slower rate than in the regular tensor product basis, rendering the nonparametric estimation of high-dimensional random coefficient models feasible.
The derived estimator allows estimating the underlying distribution with constrained least squares, making the approach computationally simple and fast.
Monte Carlo experiments and an application to data on the regulation of air pollution illustrate the good performance of the estimator.
		
		

\vspace{0.4cm}
		
 \vspace{0.1cm}\newline

\end{abstract}

\vfill
\noindent\hrulefill \\
{\footnotesize{
Computational infrastructure and support were provided by the Centre for Information and Media Technology at Heinrich Heine University Düsseldorf.\\
$\dagger$ University of Groningen, P.O. Box 800, 9700 AV Groningen, The Netherlands. \\ Email: \href{mailto:m.k.osterhaus@rug.nl}{m.k.osterhaus@rug.nl} 
}}

\end{titlingpage}

\onehalfspacing

\section{Introduction}

Adequately modeling unobserved heterogeneous behavior of economic agents is a common challenge in many empirical economic studies. Random coefficient models are frequently applied to address this challenge. They allow the coefficients of the model to vary across agents according to an unknown distribution. 
Conventional parametric estimators typically assume that the random coefficients follow a certain family of distributions up to some unknown finite-dimensional parameters. However, such estimators lack flexibility as they are often limited to a few families of distributions, and are restrictive as they rely on the assumption that the assumed distribution is correct. 
Due to the increasing availability of large data sets, nonparametric estimators for random coefficient models become more and more attractive for applied research. These estimators allow to recover distributions from the data without such limiting prior assumptions on the shape of the distribution.

A popular nonparametric approach is the method of sieves \citep{chen2007}. Sieve estimators approximate the underlying distribution using a finite number of basis functions that typically increase with the sample size. Unfortunately, sieve estimators can quickly lead to imprecise estimates or even become computationally unfeasible when the model includes multiple random coefficients. Because the standard way to extend one-dimensional basis functions to multi-dimensional functions is a tensor product construction, the number of parameters increases exponentially in the number of random coefficients. This property, known as the curse of dimensionality, limits the application of such estimators to models with only a few random coefficients -- even if the number of basis functions in one dimension is moderately small.  

This paper proposes and investigates a sparse grid approach for the nonparametric estimation of high-dimensional random coefficient models. The estimator approximates the underlying distribution using a linear combination of multi-dimensional hierarchical basis functions.  
The hierarchical structure of the basis functions has two major advantages: First, their local support makes it possible to accurately approximate the local peculiarities of the distribution without imposing certain functional forms in other regions. Second, they are particularly suited for the construction of sparse bases. 
For the construction of a sparse hierarchical basis, we adopt the sparse grid method suggested by \citet{zenger91}. 
The approach uses a truncated tensor product which reduces the number of basis functions substantially. Because smoother functions can typically be approximated by a smaller number of basis functions \citep{hansen2014}, the truncated tensor product deteriorates the approximation accuracy only slightly if the underlying random coefficients distribution is sufficiently smooth (e.g., see \citealp{bungartz2004}). 
In addition to the sparse tensor product construction, we study a spatially adaptive refinement procedure for estimating non-smooth distribution functions. Depending on the local shape of the underlying distribution, the spatially adaptive refinement incrementally adds basis functions in those areas of the distribution where the improvement in the overall approximation accuracy is highest.
To provide a computationally simple and fast estimator, we exploit linearity using the linear probability model transformation suggested by \citet{fox2011}. This way, the parameters of the model can be estimated using constrained least squares. 

We study the finite sample properties of our estimator in various Monte Carlo experiments. Using the nonparametric estimator of \citet{fox2011} as a benchmark, our estimator provides comparably accurate approximations of the true underlying distribution, even if the distribution has a steep and wiggly curvature. Moreover, the results confirm the theoretical properties of the sparse grid approach. The accuracy of the estimator slowly declines with an increasing number of random coefficients included into the model and with decreasing smoothness of the true distribution. For non-smooth distributions, the spatially adaptive refinement improves the approximation accuracy remarkably. Because the estimator becomes more accurate with increasing sample size if the number of basis functions is sufficiently large, the estimator can be viewed as a sieve estimator \citep{chen2007}. 
An application to the model of dynamic regulation of air pollution in \citet{blundell2020} emphasizes the advantage of our estimator. \citet{blundell2020} estimate the five-dimensional distribution with the fixed grid estimator of \citet{fox2011} using $10,001$ grid points. 
Even though our estimator requires substantially fewer parameters, the estimated results are similar to those of \citet{blundell2020} -- especially with respect to the estimated predictions of the conducted counterfactual experiments.


The underlying principle of sparse grids -- a sparse tensor product decomposition -- goes back to the seminal work of \citet{smolyak1963}. Sparse grids for estimating nonlinear models in economics (including random coefficient models) have been studied by \citet{heiss2008}. In contrast to our estimator, the approach of \citet{heiss2008} studies sparse grids in combination with quadrature rules for numerical integration, thereby restricting the approach to the parametric estimation of random coefficient models. Sparse grids in combination with hierarchical basis functions have been used in several research areas for function approximation and interpolation to overcome the curse of dimensionality. Among others, \citet{ma2009} employ the concept for the solution of stochastic differential equations (a frequent challenge in physics and engineering), \citet{pflueger2010_2} for high dimensional classification problems (in data mining), and \citet{bungartz2014} and \citet{pflueger_2016} for nonparametric density estimation. The only application of sparse hierarchical bases in economics which we are aware of is by \citet{brumm2017}. They employ a sparse hierarchical basis for the interpolation of macroeconomic policy functions in dynamic optimization problems.

Our sparse grid estimator primarily relates to the nonparametric fixed grid estimator of  \citet{train2008}, \citet{fox2011} and \citet{heiss2021}, and to the nonparametric estimator of \citet{train2016}. Both estimators use linear sieves to approximate the underlying random coefficients' distribution. 
The fixed grid approach uses a set of fixed support points and estimates the probability mass at every point from the data. The disadvantage of the approach is that the number of parameters equals the number of support points, leading to a large number of parameters if the estimated distribution is supposed to be smooth -- especially if the model has multiple random coefficients. In fact, \citet{fox2016} show that the fixed grid estimator suffers from the curse of dimensionality as the derived error bound of the estimated distribution function is less tight if the number of random coefficients increases.
\citet{train2016} proposes to approximate the random coefficients' distribution using polynomials, splines or step functions as basis functions inside logit kernels to model the shape of the distribution. The logit kernel assures nonnegativity of the probability mass at each support point and summation to one. The parameters of the model are estimated with simulated maximum likelihood. In order to avoid an exponential increase in the number of basis functions, \citet{train2016} proposes to use mainly one-dimensional basis functions and to include only few multi-dimensional basis functions to capture the correlation across dimensions. In contrast to the approach proposed in this paper, the approach proposed by \citet{train2016} lacks theoretical guidance on the choice of basis functions. 



The remainder of the paper is organized as follows: Section \ref{secHB:estimatorHB} presents a nonparametric estimator for random coefficients distribution using a linear combination of basis functions. Section \ref{secHB:basis} explains the construction of sparse hierarchical bases, and Section \ref{secHB:adaptive} presents a spatially adaptive refinement procedure of the sparse hierarchical basis. Section \ref{secHB:MonteCarloHB} studies the performance of the estimator in several Monte Carlo experiments, and Section \ref{secHB:ApplicationHB} presents an application to real data. Section \ref{secHB:ConclusionHB} concludes.


\section{Estimator}\label{secHB:estimatorHB}

This section briefly lays out the random coefficients model and presents a computationally simple and fast nonparametric estimator that approximates the true distribution using a linear combination of basis functions. The estimator is general in the sense that it can be applied using any type of basis functions. The construction of sparse hierarchical bases is deferred until Section \ref{secHB:estimatorHB}.

Consider the following random coefficient discrete choice model. Let there be an i.i.d. sample of $N$ observations, each confronted with a set of $J$ mutually exclusive potential outcomes, and an outside option. The researcher observes a $D$-dimensional real-valued vector of explanatory variables, $\bm{x}_{n,j}=(x_{n,j,1}, \ldots, x_{n,j,D})$, for every observation unit $n$ and potential outcome $j$, and a unit vector $\bm{y}_i$ whose entries are equal to one when she observes outcome $j$ for the $n$th observation unit, and zero otherwise.\footnote{\textit{Notation}: In the following, vectors and matrices will be written in boldface.} 
Denote the probability of outcome $j$ for a given covariate vector $\bm{x}_{n,j}$ and random coefficient vector $\bm{\beta}_n=(\beta_{n,1}, \ldots, \beta_{n,D})\in \mathbb{R}^D$ by $g\left(\bm{x}_{n,j}, \bm{\beta}_n\right)$. The functional form of $g(\cdot)$ is specified by the researcher. 
Integrating the conditional outcome probability $g\left(\bm{x}_{n,j}, \bm{\beta}_n\right)$ over the distribution of random coefficients yields the unconditional probability that outcome $j$ occurs for observation $n$ given covariates $\bm{x}_{n,j}$,
\begin{equation}\label{eqHB:uncondprob}
P_{n,j}\left(\bm{x}\right) = \int_{\Omega_1}\cdots\int_{\Omega_D} g\left(\bm{x}_{n,j}, \bm{\beta}\right)f_0\left(\bm{\beta}\right)\diff\beta_D\ldots\diff\beta_1,
\end{equation}
where $f_0(\bm{\beta}):\Omega\rightarrow\mathbb{R}_+$ represents the joint probability density function of the unknown random coefficients' distribution with domain $\Omega$.

The goal of the researcher is to estimate the unknown distribution from the data. A popular nonparametric approach for this task is the method of linear sieves. Linear sieves use a finite linear combination of prespecified basis functions (e.g., polynomials or splines) to approximate functions of unknown shapes. 
Define $\Phi_B:=\{\phi_{b}\}_{b=1}^B$ as the finite set of such basis functions with corresponding  approximation space $V_B$. The number of functions in $\Phi_B$, $B$, and the shape are specified by the researcher.\footnote{In order to select the domain of the basis, the researcher can use some preliminary estimates. For instance, estimating the distribution with a parametric approach first and then centering the grid at the mean estimates and taking multiple standard deviations to specify the domain.} 
Starting from the approximation of the true probability density function, $f_0(\bm{\beta})$, through a linear combination of the basis functions in $\Phi_B$, 
\begin{equation}\label{eqHB:approx_dens_true}
f_0\left(\bm{\beta}\right)\approx\tilde{f}(\bm{\beta}):=\sum_{b=1}^B\alpha_b\phi_b(\bm{\beta})
\end{equation}
the approximated unconditional outcome probabilities are
\begin{equation}\label{eqHB:Ptilde}
P_{n,j}\left(\bm{x}\right)\approx\tilde{P}_{n,j}\left(\bm{x}\right)=\int_{\Omega_1}\cdots\int_{\Omega_D} g\left(\bm{x}_{n,j}, \bm{\beta}\right) \sum\limits_{b=1}^B\alpha_{b}\phi_{b}\left(\bm{\beta}\right)\diff\beta_D\ldots\diff\beta_1,
\end{equation}
where $\bm{\alpha}:=(\alpha_1, \ldots, \alpha_B)\prime$ denote coefficients to be estimated. The $\approx$ arises from the approximation of the true joint probability density function $f_0(\bm{\beta})$ through $\tilde{f}(\bm{\beta})$.
For the estimation of $\bm{\alpha}$, we adopt the approach of \citet{fox2011} and transform Equation (\ref{eqHB:Ptilde}) into a linear probability model. Adding $y_{n,j}$ to both sides and moving $P_{n,j}$ to the right yields
\begin{equation}\label{eqHB:linearProbModel}
y_{n,j} \approx\sum\limits_{b=1}^B\alpha_{b}\int_{\Omega_1}\cdots\int_{\Omega_D} g\left(\bm{x}_{n,j}, \bm{\beta}\right)\phi_{b}\left(\bm{\beta}\right)\diff\beta_D\ldots\diff\beta_1 + \left(y_{n,j} - P_{n,j}\left(\bm{x}\right)\right). 
\end{equation} 
where we used the sum rule of integration, thereby restricting the summation terms Equation \eqref{eqHB:Ptilde} to be finite.
Equation (\ref{eqHB:Ptilde}) reveals two computationally desirable properties: First, the coefficients $\bm{\alpha}$ are independent of the integral, implying that the integral needs to be simulated only once prior to the estimation. 
Second, the coefficients enter the unconditional choice probabilities linearly. 
For the estimation of $\bm{\alpha}$, we simulate the integral in Equation (\ref{eqHB:linearProbModel}) using a finite set of nodes $\mathcal{B}_R:=\{\bm{\beta}_r\}_{r=1}^R$ (e.g., using Halton or Sobol quasi-random sequences),
\begin{equation}\label{eqHB:simLinearProbModel}
y_{n,j} \approx\sum\limits_{b=1}^B\alpha_{b}\sum\limits_{r=1}^R g\left(\bm{x}_{n,j}, \bm{\beta}_r\right)\phi_{b}\left(\bm{\beta}_r\right) + \left(y_{n,j} - P_{n,j}\left(\bm{x}\right)\right). 
\end{equation} 
The $\approx$ is now decomposed of the error from the approximation of $f_0(\bm{\beta})$ through $\tilde{f}(\bm{\beta})$, and the approximation error arising from the numerical simulation of the integral.\footnote{By the strong law of large numbers, the approximated integral over of the basis function converges weakly to its analytic solution such that the latter approximation error approaches zero \citep{train2009} if $R$ is sufficiently large.}  
The property that $\bm{\alpha}$ enters Equation (\ref{eqHB:simLinearProbModel}) linearly allows to estimate the coefficients with constrained least squares, which is easy to implement and computationally fast. The binary outcome vector $\bm{y}=(y_{1, 1}, \ldots, y_{1, J}, \ldots, y_{N, J})$ denotes the dependent variable and $\sum_{r=1}^Rg(\textbf{x}_{n,j}, \bm{\beta}_r)\phi_b(\bm{\beta}_r)$ the $bth$ regressor -- the regression in total has $NJ$ observations, $J$ ``regression observations'' for every statistical observation unit $n = 1, \ldots, N$ and $B$ regressors. 
In order to estimate a valid distribution function, we estimate the coefficient vector $\bm{\alpha}$ subject to the constrains that $\tilde{f}(\bm{\beta})$ is nonnegative and has unit integrand, 
\begin{equation}\label{eqHB:CLSHB}
\hat{\bm{\alpha}} = \argmin\limits_{\bm{\alpha}\in\Lambda}\frac{1}{2NJ}\sum\limits_{n = 1}^N\sum\limits_{j = 1}^J\left(y_{n,j} - \sum\limits_{b=1}^{B}\alpha_{b}\sum\limits_{r=1}^R g\left(\bm{x}_{n,j}, \bm{\beta}_r\right)\phi_{b}\left(\bm{\beta}_r\right) \right)^2 
\end{equation} 
where $\Lambda:=\{\bm{\alpha}\in\mathbb{R}^D: \sum_{b=1}^B\alpha_b\phi_b(\bm{\beta}_r)\geq 0 \ \forall \ \bm{\beta}_r\in\mathcal{B}_R, \sum_{b=1}^B\alpha_b\sum_{r=1}^R\phi_b(\bm{\beta}_r)=1\}$.\footnote{The estimator relates to the smooth basis densities estimator proposed in \citet{fox2011}. They propose to approximate the true distribution through a mixture of normal densities, and estimate the probability weight of each normal subject to the constraint that the weights are nonnegative and sum to one. The proposed estimator is a special case of nonnegative lasso (see \citet{heiss2021} for more details), leading to sparse solutions. In contrast, our estimator does not relate to the lasso and, hence, does not suffer from sparsity.} 
By the definition of choice probabilities, the expected value of the composite error term $y_{n,j} - P_{n,j}(\bm{x}_{n,j})$ conditional on $\bm{x}_{n,j}$ is zero, such that the regression model satisfies the mean-independence assumption of the least squares approach \citep{fox2011}.
The optimization problem stated in Equation (\ref{eqHB:CLSHB}) is convex and has a single global optimum if the basis functions in $\Phi_B$ are linearly independent. It can be solved with common statistic software using specialized optimization routines (e.g., R's solve.QP function from the quadprog package or MATLAB's lsqlin function).\\

The estimated joint distribution function at point $\bm{\beta}$ is constructed from the weighted sum of the estimated coefficients and basis functions,
\begin{equation}\label{eqHB:approxCDF}
\hat{F}\left(\bm{\beta}\right) = \sum\limits_{b=1}^B\hat{\alpha}_{b}\sum\limits_{r=1}^R 1\left[\bm{\beta}_r\leq\bm{\beta}\right]\phi_{b}\left(\bm{\beta}_r\right),
\end{equation}
where $1[\bm{\beta}_r\leq\bm{\beta}]$ is an indicator function that is equal to one whenever $\bm{\beta}_r\leq\bm{\beta}$, and zero otherwise. The term to the right of coefficient $\alpha_b$ corresponds to the simulated integral of the corresponding basis function $\phi_b(\cdot)$ with upper bound $\bm{\beta}$ using $R$ simulation nodes. 
The estimated distribution approximates the true underlying distribution through a discrete distribution with $R$ support points and probability weight $\hat{f}(\bm{\beta}_r)=\sum_{b=1}^B\hat{\alpha}_b\phi_b(\bm{\beta}_r)$ at every point $r=1,\ldots,R$.

For the estimation of multi-dimensional random coefficients distributions, the multi-dimensional bases are typically constructed using a regular tensor product of one-dimensional basis functions. Starting from a one-dimensional basis with $B$ basis functions, the $D$-dimensional regular tensor product basis includes $B^D$ basis functions \citep{chen2007}. Because the exponential dependency renders the approach computationally unfeasible for high-dimensional distributions, the above estimator with a regular tensor basis is limited to moderately low-dimensional random coefficient models.

\begin{remark}
The proposed estimator can be easily extended to a generalized least-squares version and a simulated maximum likelihood version. For the generalized least-squares version, each ``regression observation'' in Equation (\ref{eqHB:CLSHB}) is weighted by a weighting matrix to address the heteroscedasticity problem associated with linear probability models and the correlation across observations that belong to the same observation unit $n$. For a detailed description of the calculation of an efficient weighting matrix, see \citet{fox2011}.

As an alternative to constrained least squares, the coefficients $\bm{\alpha}$ can be estimated with simulated maximum likelihood using the approach of \citet{train2016}, who proposes to model the probability weight at support point $\beta_r$ using a linear combination of basis functions inside a logit kernel. The exponential function in the logit kernel assures that the estimated weights are positive. The denominator normalizes the probability weights such that they sum up to one.
\end{remark}

\begin{remark}
When choosing the family of basis functions and the number of simulation draws $R$, it is important that the basis functions are linearly independent, and that the researcher chooses $R$ to be sufficiently large such that the draws are sufficient to cover the domain densely. 
If the number of simulation draws is too small, there are only a few simulation draws inside the support of every basis functions with the consequence that most column entries are zero. This property can lead to an ill-conditioning of the least squares problem \citep{judd2011}.\footnote{In addition to the multicollinearity problem, choosing fewer simulation draws can also lead to poor scaling of the regressor matrix. If there are only a few simulation draws inside the support of every basis function, the columns of the regressor matrix have only a few very small entries, in which case they are treated as if they are columns of zeros \citep{judd2011}.}
One tool recommended in the literature to improve the numerical stability of least squares problems is Thikonov regularization \citep{hoerl1970} (see, e.g., \citealp{judd2011}, \citealp{cohen2013}, or \citealp{pflueger2010_2}), which is already successfully used by \citet{heiss2021} to improve the performance of the nonparametric fixed grid estimator of \citet{fox2011}.\footnote{We noticed that the instability of the estimator for high levels when the distribution is estimated with ordinary least squares disappears when the coefficients are estimated with constrained least squares. To this end, the constraints seem to constitute a form of regularization that stabilizes the estimator and potentially makes additional regularization redundant.} 
\end{remark}

\section{Sparse Hierarchical Bases}\label{secHB:basis}

This section explains the construction of sparse hierarchical bases. Because the sparse grid is based on a truncated tensor product of one-dimensional hierarchical basis functions, we start with the concept of hierarchical basis functions, and then explain how sparse grids can be constructed from multi-dimensional hierarchical bases. For a more comprehensive presentation of hierarchical bases and sparse grids, see, e.g., \citet{bungartz2004} and \citet{garcke2012sparse}.

\subsection{Hierarchical Multilevel Bases} \label{subsecHB:hierbasis_functions}

Hierarchical bases are based on a decomposition of the approximation space into a finite number of hierarchically structured segments -- intervals in the univariate case and hyper-rectangles in the multivariate case. These segments are constructed via a discretization of the domain $\Omega$ of the function under consideration using  equidistant grids. 
In the following, we consider the $D$-dimensional unit cube, $\Omega=[0, 1]^D$ for ease of notation. The construction of the hierarchical basis can be easily adapted to different domains via rescaling. Furthermore, we assume that $f_0$ is vanishing on the boundary of $\Omega$ ($f_0\vert_{\delta\Omega}=0$).\footnote{The restriction that $f_0$ is vanishing at the boundary of $\Omega$ can be overcome by adding basis functions that are nonzero at the boundaries (for more details, see, e.g., \citet{pflueger2010}). \citet{train2016} points out that it can be beneficial to restrict the function to be zero at the boundaries of the domain as this eliminates the long tails of some distributions, e.g., of the normal or lognormal distribution, which can be unrealistic in real-world applications.}

Let $l\in\mathbb{N}$ denote the discretization level specified by the researcher. 
In the one-dimensional case, the grid $\Omega_l$ with points $b_{l,i}:=2^{-l}\ i$ and mesh size $h_l:=2^{-l}$ splits the domain $\Omega$ into $2^l$ equally-sized intervals. The index $i\in\mathbb{N}$ indicates the location of a grid point. 
Every grid point is associated with a basis function $\phi:[0, 1]\rightarrow\mathbb{R}$ that is centered at the corresponding grid point. For the construction of the sparse grid basis, we consider the piecewise-linear hat function
\begin{equation}
\phi(\beta) :=
    \begin{cases}
      1 - \left\vert\beta\right\vert, & \text{if $\beta\in\left[-1, 1\right]$}\\
      0, & \text{otherwise.}
    \end{cases} 
\end{equation}
Using translation and scaling according to level $l$ and index $i$, the basis function centered at grid point $b_{l,i}$ is 
\begin{equation}\label{eqHB:bfunction}
\phi_{l,i}\left(\beta\right) := \phi\left(\frac{\beta - b_{l,i}}{h_l}\right)
\end{equation}
with $\phi_{l,i}(b_{l,i}) = 1$ and local support $[b_{l,i} - h_l, b_{l,i} + h_l]$. 

\begin{figure}[h]%
\captionsetup{justification=centering, font=normalsize, skip=10pt} 
\caption{One-Dimensional Piecewise-linear Hierarchical Basis Functions}
\includegraphics[width=1\textwidth, scale=1]{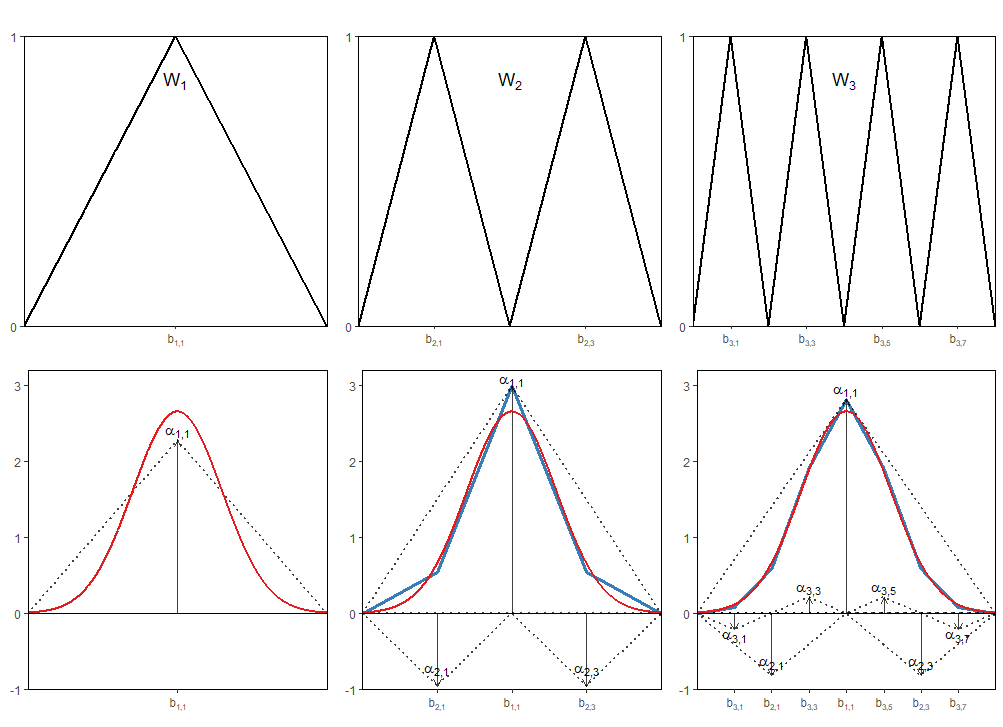}
\captionsetup{justification=justified,singlelinecheck=true, width=1\textwidth, font={footnotesize,stretch=0.8}, skip=2pt} 
\caption*{\textit{Note:} The top panel shows the one-dimensional hierarchical subspaces $W_l$ for $l=1$ (left), $l=2$ (center), and $l=3$ (right). The bottom panel illustrates the approximation of a univariate normal density (solid red line) with mean $0.5$ and standard deviation $0.15$, $f(\beta)=\phi(\beta\vert 0.5, 0.15^2)$, through a one-dimensional piecewise-linear hierarchical basis $\Phi_l$ with levels $1, 2, 3$ (solid blue line). The contribution of every basis function to the approximation is indicated by the grey arrows.}
\label{fig:hierBasis1d}
\end{figure}

To construct a basis with hierarchically arranged functions, the locations of the grid points -- and the number of basis functions within a level -- are determined by the index sets 
\begin{equation}\label{eqHB:1DIndex}
\mathcal{I}_l:=\left\{i\in\mathbb{N}: \ 1\leq i \leq 2^l-1, i \ \text{odd}\right\}. 
\end{equation} 
All basis functions with level $l$ centered at the grid points corresponding to index set $\mathcal{I}_l$ span the hierarchical subspace $W_l$,
\begin{equation}\label{eqHB:subspaces1D}
W_l := \spanS\left\{\phi_{l, i} : i\in \mathcal{I}_l\right\}.
\end{equation} 
The upper panel in Figure \ref{fig:hierBasis1d} illustrates the one-dimensional piecewise-linear hierarchical subspaces $W_l$ going from level $l=1$ (left) to level $l=3$ (right). 
\captionsetup{justification=justified, singlelinecheck=false}
All hierarchical basis functions with the same level have the same size, shape and compact support. While the number of basis functions that span a subspace increase with the level $l$ of the subspace, the support of each function decreases with $l$. The index sets $\mathcal{I}_l$ ensure that (i) different basis functions within the same level have mutually disjoint support, and (ii) that the support of a basis function with level $l$ nests the support of two basis functions of the next higher level, $l+1$. 

The hierarchical basis of level $l$ is the set of all basis functions with level $1\leq k\leq l$ and corresponding index $i\in \mathcal{I}_k$, 
\begin{equation}
\Phi_l := \left\{\phi_{k,i}: \ i\in \mathcal{I}_k, 1\leq k \leq l\right\}.
\end{equation}
The bottom panel in Figure \ref{fig:hierBasis1d} illustrates the approximation of a univariate normal density using a one-dimensional piecewise-linear hierarchical basis. Due to the hierarchical structure, hierarchical bases of different levels are nested such that the basis of a level $l$ refines a basis of the next lower level. The smaller support of basis functions with a higher level allows to approximate local peculiarities more accurately. 
The shape of the approximated function depends on the shape of the specified type of basis function. For instance, the one-dimensional piecewise-linear hierarchical basis approximates the true probability density function on every segment by a linear function.\\

Starting from the one-dimensional hierarchical basis, a $D$-dimensional hierarchical basis on $\Omega=[0, 1]^D$ is obtained via a tensor product construction. Let the multi-index $\bm{l}=(l_1, \ldots, l_D)\in\mathbb{N}^D$ denote the discretization level of the hierarchical basis in every dimension, and $\bm{i}\in\mathbb{N}^D$ indicate the spatial position of the $D$-dimensional grid points. In the following, all relational operations involving vectors are to be read component-wise.
The $D$-dimensional grid $\Omega_{\bm{l}}$ with grid points $\bm{b_{l,i}}:=(b_{l_1,i_1}, \ldots, b_{l_D,i_D})$ and mesh size $\bm{h_l}=(h_ {l_1}, \ldots, h_{l_D})$ can be constructed from the cartesian product of one-dimensional grids in every dimension. Accordingly, the indices $i_d$ and $l_d$ can vary across $d$ for a given grid point. The grid points are equidistant in each dimension but can differ across dimensions (e.g., the grid can be finer in more important dimensions). 

\begin{figure}[h]
\captionsetup{justification=centering, font=normalsize, skip=10pt} 
\caption{Two-Dimensional Piecewise-bilinear Hierarchical Basis Functions.}
    \centering
    \subfigure{\includegraphics[width=0.49\textwidth]{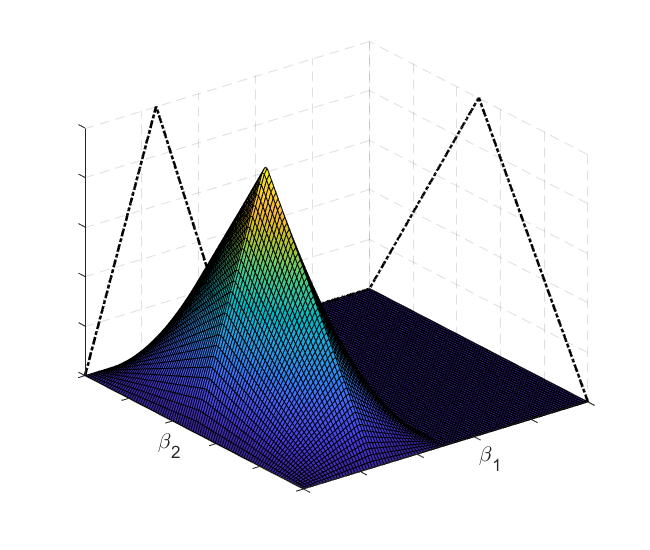}} 
    \subfigure{\includegraphics[width=0.49\textwidth]{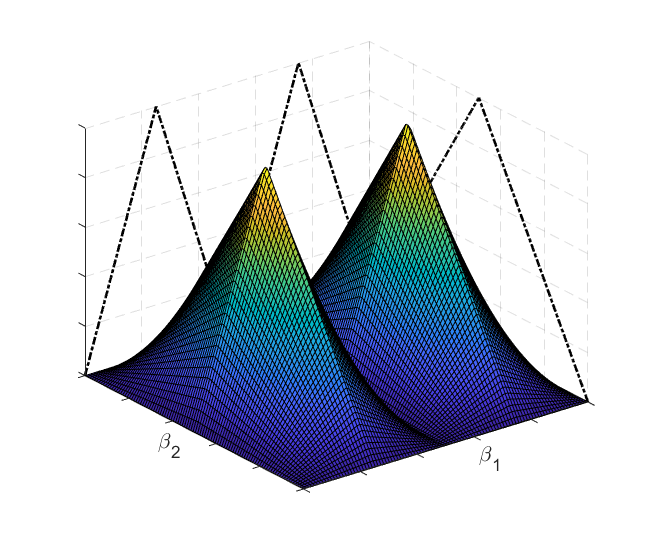}}
    \captionsetup{justification=justified,singlelinecheck=true, width=1\textwidth, font={footnotesize,stretch=0.8}, skip=2pt} 
\caption*{\textit{Note:} The left panel illustrates the tensor product construction of the two-dimensional piecewise-bilinear hierarchical basis function for level $\bm{l}=(2, 1)$ and index $\bm{i}=(1, 1)$ from the one-dimensional piecewise-linear basis function $\phi_{2, 1}$ in dimension $d=1$ and $\phi_{1, 1}$ in dimension $d=2$ (dashed black lines). The right panel shows the hierarchical subspace $W_{(2, 1)}=\{\phi_{(2, 1), (1, 1)}, \phi_{(2, 1), (3, 1)}\}$.}
    \label{fig:tensor}
\end{figure}

As in the one-dimensional case, every grid point spans a basis function with support on the respective segment. The $D$-dimensional basis function centered at grid point $\bm{b_{l,i}}$ is defined as the product of one-dimensional basis functions,
\begin{equation}
\phi_{\bm{l},\bm{i}}\left(\bm{\beta}\right):=\prod\limits_{d=1}^D\phi_{l_d, i_d}\left(\beta_d\right).
\end{equation} 
The left panel in Figure \ref{fig:tensor} illustrates the tensor product construction of a two-dimensional piecewise-bilinear hierarchical basis function with level $\bm{l}=(2, 1)$ and index $\bm{i}=(1, 1)$ from the one-dimensional piecewise-linear basis function $\phi_{2, 1}$ in dimension $d=1$ and $\phi_{1, 1}$ in dimension $d=2$ (dashed black lines).

The multivariate hierarchical subspaces are defined analogously to the univariate case,
\begin{equation}\label{eqHB:subspacesMultiD}
W_{\bm{l}}:=\spanS\left\{\phi_{\bm{l},\bm{i}}:\ \bm{i}\in \mathcal{I}_{\bm{l}}\right\}, \quad\mathcal{I}_{\bm{l}}:=\mathcal{I}_{l_1}\times\cdots\times\mathcal{I}_{l_D},
\end{equation}
where the $D$-dimensional index sets $\mathcal{I}_{\bm{l}}$ can be constructed as the cartesian product of the one-dimensional index sets.
The right panel in Figure \ref{fig:tensor} shows the hierarchical subspace $W_{(2, 1)}$ which is spanned by the basis functions $\phi_{(2, 1), (1, 1)}$ and $\phi_{(2, 1), (3, 1)}$.

The $D$-dimensional hierarchical basis is the set of all basis functions with level $\bm{1}\leq\bm{k}\leq \bm{l}$ and index $\bm{i}\in \mathcal{I}_{\bm{k}}$
\begin{equation}
\Phi_{\bm{l}}:=\left\{\phi_{\bm{k},\bm{i}}: \bm{i}\in \mathcal{I}_{\bm{k}}, \bm{1}\leq\bm{k}\leq \bm{l} \right\}.
\end{equation}
The approximated function $\tilde{f}_{\bm{l}}\in V_{\bm{l}}$ is a linear combination of $D$-dimensional hierarchical basis functions with coefficients $\alpha_{\bm{k},\bm{i}}\in\mathbb{R}$,
\begin{equation}\label{eqHB:multiHB}
f_0\left(\bm{\beta}\right)\approx\tilde{f}_{\bm{l}}\left(\bm{\beta}\right) := \sum\limits_{\bm{k}=\bm{1}}^{\bm{l}}\sum\limits_{\bm{i}\in I_{\bm{k}}}\alpha_{\bm{k},\bm{i}}\phi_{\bm{k},\bm{i}}\left(\bm{\beta}\right)
\end{equation} 
where $V_{\bm{l}}$ denotes the function space spanned by the hierarchical basis functions.
Due to the linear independence across piecewise-linear hierarchical basis functions, the underlying function $f_0$ can be uniquely approximated through $\tilde{f}_{\bm{l}}$ \citep{valentin2016}. 

\citet{bungartz2004} show that the full hierarchical basis includes $\vert V_{\bm{l}}\vert=(2^l-1)^D=\mathcal{O}(2^{lD})$ basis functions. 
The full hierarchical basis has the same limitation as other existing nonparametric estimators for random coefficient models. Due to the regular tensor product construction of multi-dimensional basis functions from one-dimensional basis functions, the number of parameters increases exponentially in the number of random coefficients, prohibiting an accurate approximation of the underlying distribution if the model includes multiple random coefficients. For instance, for a $5$-dimensional distribution and level $l=3$ in each dimension, estimating the model using the full hierarchical basis involves the estimation of $(2^3 - 1)^5=16,807$ parameters.

\subsection{Classical Sparse Grids}\label{subsecHB:sparse}

To alleviate the curse of dimensionality, sparse grids seek to construct an approximation space that is better than the full grid space $V_{\bm{l}}$ in the sense that the same number of basis functions leads to a higher approximation accuracy. 
The classical sparse grid approach \citep{zenger91} takes advantage of the hierarchical nature of the basis functions. 
Starting from the definition of the approximation space $V_{\bm{l}}$ as a direct sum of hierarchical subspaces,\footnote{The space $V_{\bm{l}}$ is the direct sum of subspaces $W_{\bm{k}}$, $\bm{1}\leq\bm{k}\leq\bm{l}$, if $V_{\bm{l}}=W_{\bm{1}} + \ldots + W_{\bm{k}}$ and the subspaces $\{W_{\bm{1}}, \ldots, W_{\bm{l}}\}$ are disjoint \cite[p. 48]{gentle2007}.}
\begin{equation}\label{eqHB:V_full}
V_{\bm{l}}:=\bigoplus\limits_{\bm{k}\leq\bm{l}}W_{\bm{k}},
\end{equation}
the approach reduces the number of basis functions by selecting only those subspaces that contribute most to an accurate approximation. The selection of subspaces arises from a discrete optimization that weights the approximation benefit of a hierarchical subspace - measured in terms of its contribution to the overall approximation in the $L_2$ norm - against its cost - measured in terms of the number of parameters \citep{bungartz2004}. 

The cost of subspace $W_{\bm{l}}$ can be immediately derived from its corresponding index set $\mathcal{I}_{\bm{l}}$, and is given by $\vert W_{\bm{l}}\vert = \vert \mathcal{I}_{\bm{l}}\vert =2^{\vert\bm{l}-\bm{1}\vert_1}$.
The contribution of a subspace to the approximation accuracy depends on the smoothness of the function under consideration, or more precisely, on its function class. The Classical sparse grid is derived for functions that are assumed to be sufficiently smooth, i.e., with bounded second-order mixed derivatives. This function class belongs to the mixed Sobolov space (of functions vanishing on the boundary)\footnote{Assuming a certain smoothness class of functions often required in the nonparametric estimation literature when studying the approximation accuracy of estimators (see, e.g., \citet{chen2007} for different smoothness classes). For example, when deriving the error bound for  the nonparametric fixed grid estimator of \citet{fox2011}, \citet{fox2016} assume that the density of the true underlying random coefficients distribution is a $s$-times continuously differentiable density function with all own and partial derivatives uniformly bounded (with respect to the $L_2$-norm) by a constant $\bar{C}<\infty$. Their derived error bound on the true and estimated distribution is tighter if the true probability density function is smoother.}
\begin{equation}
\mathcal{H}^2_{\text{mix}}\left(\Omega\right):=\Big\{f:\Omega\rightarrow\mathbb{R}:D^{\bm{r}}f\in L_2(\Omega), \vert\bm{r}\vert_\infty\leq c, f\vert_{\partial\Omega}=0\Big\}
\end{equation}
with $\vert \bm{r}\vert_\infty:=\max_{1\leq d\leq D}r_d$ and smoothness parameter $c=2$, where $\mathcal{D}^{\bm{r}}$ denotes the differential operator defined by  
\begin{equation}
\mathcal{D}^{\bm{r}}:=\frac{\partial^{\bm{r}}}{\partial\beta_1^{r_1}\cdots\partial\beta_D^{r_D}}
\end{equation}
given a $D$-tuple $\bm{r}=(r_1, \ldots, r_D)$ of nonnegative integers.
Recall that in the representation of the probability density function as a weighted sum of hierarchical basis functions, the coefficient $\alpha_{\bm{k},\bm{i}}$ indicates the refinement of the local approximation constructed with those functions of the next lower level, $\bm{l-1}$, through the function with level $\bm{l}$. \citet{bungartz2004} show for functions $f\in \mathcal{H}^2_{\text{mix}}(\Omega)$ that the coefficients in the representation of the underlying function as a linear combination of piecewise-linear hierarchical basis functions decay rapidly as the level increases,
\begin{equation}
\vert \alpha_{\bm{l},\bm{i}}\vert = \mathcal{O}\left(2^{-2\vert \bm{l}\vert_1}\right),
\end{equation}
where $\vert \bm{l}\vert_1:=\sum_{d=1}^D l_d$. 
Thus, the decreasing support of basis functions with increasing level together with the decay of the coefficients imply a decreasing contribution of subspaces with higher level if the underlying function is sufficiently smooth. 

\begin{figure}[h!]
\captionsetup{justification=centering, font=normalsize, skip=10pt} 
\caption{Two-Dimensional Full Grid, Sparse Grid, and Spatially Adaptive Sparse Grid for Level $l=3$.}
\setlength\belowcaptionskip{6pt}
    \centering
    \includegraphics[width=1\textwidth]{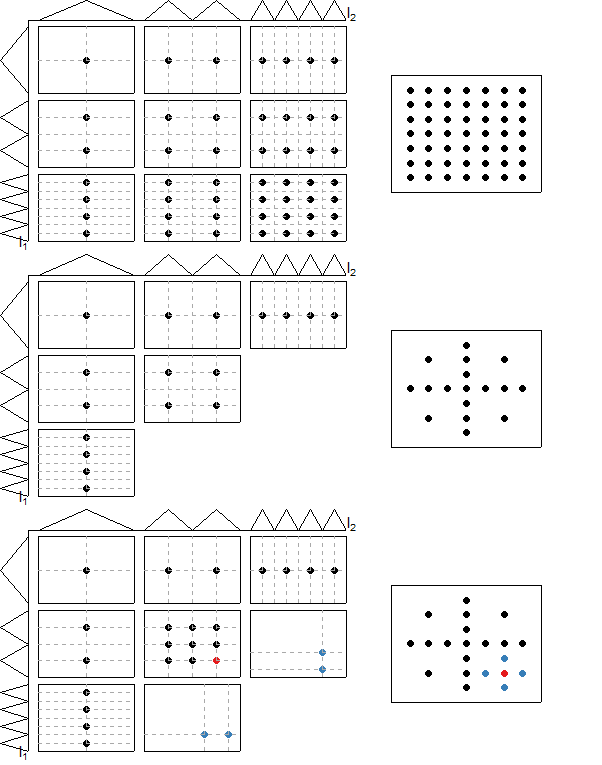}
    \captionsetup{justification=justified,singlelinecheck=true, width=1\textwidth, font={footnotesize,stretch=0.8}, skip=2pt} 
\caption*{\textit{Note:} The upper panel illustrates the construction of a two-dimensional full Cartesian grid of level $\bm{l}=(3, 3)$. The panel in the center shows the construction of a classical sparse grid of level $l_S=3$. The bottom panel illustrates the spatially adaptive refinement of the sparse grid of level $l_S=3$. The red point represents the grid point that is selected for refinement, and the blue points represent the grid points that are added to the initial sparse grid.}
    \label{fig:grids}
\end{figure}

The classical sparse grid leaves out those subspaces within the full grid space $V_{\bm{l}}$ that contribute only little to the function approximation, i.e., for which the absolute values of the coefficients are small. This is done via an a priori optimization which minimizes the approximation error (measured by the $L_2$ norm) while keeping the number of grid points fixed. For functions in the mixed Sobolov space $\mathcal{H}_2^{\text{mix}}(\Omega)$, this yields the classical sparse grid space 
\begin{equation}\label{eqHB:V_sparse}
V_{l_S}:=\bigoplus\limits_{\vert\bm{k}\vert_1\leq l_S+D-1}W_{\bm{k}}.
\end{equation} 
of level $l_S\in\mathbb{N}$ specified by the researcher.
Note that the set of basis functions that spans $V_{l_S}$ is now decomposed of only those functions from subspaces for which $\vert \bm{k}\vert_1\leq l_S + D - 1$ for every $\bm{1}\leq\bm{k}\leq\bm{l}$, 
\begin{equation}
\Phi_{l_S}:=\left\{\phi_{\bm{k},\bm{i}}:  \bm{i}\in\mathcal{I}_{\bm{k}}, \vert \bm{k}\vert_1 \leq l_S+D-1 \right\}.
\end{equation} 
Thus, the sparse grid approximation space $V_{l_S}$ is a subset of the full grid approximation space with discretization level $l_S$ in every dimension, $V_{l_S}\subset V_{\bm{l}}$ with $\bm{l}=(l_S, \ldots, \l_S)$.
Figure \ref{fig:grids} illustrates the construction of a two-dimensional classical sparse grid of level $l_S=3$. 
The number of basis functions in the sparse hierarchical basis is 
\begin{equation}\label{eqHB:n_grid_points_sparse}
\vert V_{l_S}\vert =\sum\limits_{i=0}^{l_S-1}2^i\cdot\begin{pmatrix}D-1+i \\ D-1\end{pmatrix} = 2^{l_S}\left(\frac{l_S^{D-1}}{(D - 1)!} + \mathcal{O}\left(l_S^{D-2}\right)\right).
\end{equation}
Expressed in terms of the number of basis functions in one dimension, $B=2^l-1$, Equation \ref{eqHB:n_grid_points_sparse} implies that the function space spanned by the sparse hierarchical basis is of order $\mathcal{O}(B\log(B)^{D-1})$, compared to $\mathcal{O}(B^D)$ for regular tensor product bases (\citealp{bungartz2004}, \citealp{brumm2017}).
Table \ref{tabHB:SparseGrid} reports the number of basis functions in a sparse hierarchical basis for different dimensions and discretization levels and for a regular tensor product basis with $B=(3,5,7)$ basis functions in one dimension. Clearly, the estimation of the model with a tensor product basis rapidly becomes computationally not feasible in five- and higher-dimensional problems for more than three basis functions in one dimension. The sparse grid approach renders the estimation of the corresponding random coefficients' distributions with such discretization levels computationally feasible. 
\begin{table}[h]
 \centering
\small
\tabcolsep=0.4cm
\captionsetup{justification=centering, font=normalsize} 
\caption{Number of Grid Points in a Full Cartesian Grid vs. Number of Grid Points in a Sparse Grid.}  
 \label{tabHB:SparseGrid}
\begin{tabular}{crrrrrrrrr}  
\toprule \toprule\noalign{\smallskip}  
 & \multicolumn{3}{c}{Sparse Grid} & \multicolumn{3}{c}{Tensor Product Basis}\\[0.5ex]
\cmidrule(l{0.5pt}r{9pt}){2-4} \cmidrule(l{0.5pt}r{9pt}){5-7} 
Dimension & $\vert V_2\vert$ & $\vert V_3\vert$ & $\vert V_4\vert$ & $B=3$ & $B=5$ & $B=7$ \\[0.5ex]
 \hline\\[-0.75ex]
 2 & 5 & 17 & 49 & 9 & 25 & 49 \\
 3 & 7 & 31 & 111 & 27 & 125 & 343 \\
 4 & 9 & 49 & 209 & 81 & 625 & 2401\\
 5 & 11 & 71 & 351  & 243 & 3125 & 16807 \\ 
 6 & 13 & 97 & 545 & 729 & 15,625 & 117,649\\ 
 8 & 17 & 161 & 1,121 & 6561 & 390,625 & $5.76\cdot 10^6$ \\ 
10 & 21 & 241 & 2,001 & 59,049 & $9,77\cdot 10^6$ & $2.82\cdot 10^8$ \\ 
\bottomrule\bottomrule 
\end{tabular} 
\end{table} 

Analogously to the full hierarchical basis, the approximated function $\tilde{f}_{l_S}\in V_{l_S}$ corresponds to the finite weighted sum of hierarchical basis functions centered at the sparse grid points,
\begin{equation}\label{eqHB:ftilde_sparse}
f_0\left(\bm{\beta}\right)\approx\tilde{f}_{l_S}\left(\bm{\beta}\right)=\sum\limits_{\vert\bm{k}\vert_1\leq l_S+D-1}\sum\limits_{\bm{i}\in I_{\bm{k}}}\alpha_{\bm{i},\bm{k}}\phi_{\bm{i},\bm{k}}\left(\bm{\beta}\right).
\end{equation} 
\citet{bungartz2004} show that the approximation accuracy of functions constructed with the sparse piecewise-linear hierarchical basis deteriorates only slightly from $\mathcal{O}(2^{-2\bm{l}})$ for the full hierarchical basis to $\mathcal{O}(2^{-2l_S}\cdot l_S^{D-1})$ if the function under consideration is sufficiently smooth. \\


\begin{remark}
The construction of sparse grids is not restricted to piecewise-linear functions but can be constructed using several different types of basis functions. For instance, \citet{valentin2016} consider cardinal B-Splines. If the spline functions are of odd degree, the knots of the cardinal B-Spline basis coincide with the grid points of the hierarchical basis. In fact, the hat function corresponds to the cardinal B-Spline of degree one. 
For more information on alternative basis functions, see, e.g., \citet{bungartz2004} and  \citet{pflueger2010}, respectively.
\end{remark}

\section{Spatially Adaptive Refinement}\label{secHB:adaptive}


The classical sparse grid contains those basis functions that are optimal in the sense that they deteriorate the approximation accuracy only slightly if the function to be approximated has bounded second-order mixed derivatives. For functions outside this class, i.e., functions with a wigglier and steeper curvature, spatially adaptive refinement can be used to further increase the approximation accuracy. Starting from the sparse grid, the refinement procedure incrementally adds basis functions to subregions where the underlying distribution is characterized by a steep curvature \citep{pflueger2010}.

\begin{figure}[H]
\captionsetup{justification=centering, font=normalsize, skip=10pt} 
\caption{One-dimensional Tree-like Structure of a Full Grid for Hierarchical Levels $l_S=1$ to $l_S=3$.}
\centering
\includegraphics[width=0.8\textwidth]{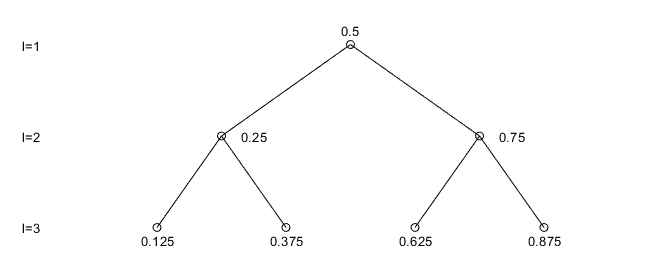}
\label{fig:tree}
\end{figure}

Due to the nested support of hierarchical basis functions of different levels, hierarchical bases are particularly suited for spatially adaptive refinement procedures. Recall that the support of a basis function of level $l-1$ is subdivided among the basis functions of the next finer level $l$ (for each function, multiple functions of the next finer level exist). Thus, by adding additional basis functions of the next finer level, one can refine the basis in some regions without affecting the approximation accuracy in others.
Figure \ref{fig:tree} illustrates this tree-like structure of a one-dimensional hierarchical basis of level $l=3$. Each grid point in the tree always serves as origin for two new points. Suppose that the point $0.25$, a point of level $l=2$, is selected for refinement. The spatially adaptive approach adds the two neighboring grid points at location $0.125$ and $0.375$ of the next higher level, $l=3$. These newly added points span basis functions with disjoint support that cover half of the support of the basis function spanned at $0.25$ (cf. Section \ref{secHB:basis}). 
For the adaptive refinement of a $D$-dimensional hierarchical basis, in every dimension, all neighboring points of the next higher level that are not yet included into the grid are added, thus keeping the coordinates of the refined grid point in the remaining dimensions fixed. This way, at most $2D$ points are added to the current grid for every point refined. The bottom panel in Figure \ref{fig:grids} illustrates the spatially adaptive procedure for a two-dimensional sparse grid of level $l_S=3$. The red point is the grid point selected for refinement, and the four blue points are those added to the current grid. The basis functions spanned at the newly added points extend the current sparse basis.

Refinable in the hierarchical structures are only grid points for which at least one neighboring point of the next higher level does not exist yet in any of the dimensions. To keep the hierarchical structure of the basis consistent, all originating points of the new grid that are not yet included have to be added if not already included. This can lead to a scenario where more than $2D$ points are added per refinement step \citep{pflueger2010}. \\

The challenge of the spatially adaptive refinement is to select those grid points for refinement that lead to the largest improvement in the approximation accuracy. A possible but computationally intensive and inefficient strategy is to consider every current grid point for refinement separately, and add only those candidates that contribute most to the problem's solution (according to a suitable error measure).\footnote{The strategy relates to the concept of backward deletion in optimal knot search for spline functions (e.g., see \citealp{wand2000}).} However, already in the one-dimensional case, there are as many grid points to be considered on the next higher level as potential new grid points as there are in the current grid. The model has to be estimated for each of these points, of which many are  unlikely to be relevant \citep{pflueger2010}.

A more efficient strategy is the identification of refinement candidates based on information available on the current grid. 
A refinement criterion commonly used in applications is the absolute value of the estimated coefficients $\alpha_{\bm{k},\bm{i}}$ (see, e.g., \citealp{pflueger2010_2} and \citealp{brumm2017}). Recall that in the reconstruction of the underlying function $f_0$ as weighted sum of hierarchical basis functions, the coefficient $\alpha_{\bm{k},\bm{i}}$ represents the local variation of $f_0$ at the area around the corresponding grid point $\bm{b}_{\bm{k},\bm{i}}$. Accordingly, refining grid points with the largest absolute value of the corresponding coefficient first promotes the refinement of those regions where the local variation of $f_0$ is strong \citep{pflueger2010}.\footnote{\citet{bungartz2014} suggest to weight the absolute value of the estimated hierarchical coefficients $\hat{\alpha}_{\bm{k},\bm{i}}$, $\vert\bm{k}\vert_1\leq l_S+D-1$ and $\bm{i}\in\mathcal{I}_{\bm{k}}$, by the function value of the basis function.} 
Another criterion suggested by \citet{pflueger2010} specifically for regression tasks is the contribution of each basis function to the squared estimated local error. Let 
\begin{equation}
\hat{\epsilon}_{n,j}^2 := (y_{n,j} - \sum_{b=1}^B\hat{\alpha}_b\sum_{r=1}^Rg(\bm{x}_{n,j}, \bm{\beta}_r)\phi_b(\bm{\beta}_r))^2\notag
\end{equation} 
denote the squared estimated local error for observation unit $n$ and alternative $j$ following from the regression in \eqref{eqHB:CLSHB}. The grid point that centers the basis function with the largest contribution to the squared local error, 
\begin{equation}
c_{\bm{l},\bm{i}}:=\sum_{n=1}^N\sum_{j=1}^J\vert \hat{\alpha}_{\bm{l},\bm{i}}\sum_{r=1}^Rg(\bm{x}_{n,j}, \bm{\beta}_r)\phi_{\bm{l},\bm{i}}(\bm{\beta}_r)\hat{\epsilon}_{n,j}^2\vert, \notag
\end{equation}
is refined first, where $R$ corresponds to the number of simulation draws,and $\hat{\bm{\alpha}}:=(\hat{\alpha}_1,\ldots,\hat{\alpha}_B)\prime$ to the vector of estimated coefficients (the estimation of the coefficients is explained in Section \ref{secHB:estimatorHB}).
We employ the criterion in the Monte Carlo experiments presented in the subsequent section, where the spatially adaptive refinement substantially improves the approximation accuracy of the sparse hierarchical basis estimator.

In addition to the refinement criterion, the researcher has to specify the number of refinement steps alongside the number of grid points refined per step. The choice depends on the problem and the data at hand. On the one hand, refining more than one grid point at once leads to a broader refinement, which can help to circumvent the refinement from getting stuck in a single characteristic of the underlying function. On the other hand, refining too many points at once expedites the increase in the number of points (especially in cases of high dimensional problems) which can cause overfitting (especially for small data sets) \citep{pflueger2010}. 

Selecting the number of points and the refinement steps relates to a model selection task. For sieve series models, \citet{hansen2014} presents a variety of different model selection procedures, the most prominent being the Akaike information criterion (AIC) and cross-validation.\footnote{Selecting the number of grid points and total refinement steps relate to the selection of the number of knots in spline regression. Another model selector than the AIC and cross-validation that is commonly used in this literature is generalized cross-validation  (e.g., see \citealp{zhou2001}, \citealp{ruppertbook}). For the spatially adaptive refinement of sparse grids, \citet{pflueger2010} suggests using $k$-fold cross-validation and to refine the hierarchical basis as long as the out-of-sample fit decreases. This approach is also used in spline regression where it is known as the myopic algorithm (the out-of-sample fit is typically measured via generalized cross-validation). We employ a full-search algorithm (\citealp[pp.~127-128]{ruppertbook}) in our Monte Carlo experiments, which calculates the out-of-sample fit for every refined grid and then selects the model with the lowest out-of-sample fit.}
While cross-validation techniques have the advantage that they take the out-of-sample fit into account to avoid over-fitting, the AIC is computationally less expensive, which is an advantage in high-dimensional problems. 
In the Monte Carlo experiments presented in Section \ref{secHB:MonteCarloHB}, we studied both $k$-fold cross-validation and the AIC for the selection of the number of refinement steps. The results indicate that both AIC and $k$-fold cross-validation appear to be suitable criteria leading to an improved approximation accuracy of the sparse hierarchical basis estimator when the local squared error is used for the selection of the grid points to be refined.




\section{Monte Carlo Simulations}\label{secHB:MonteCarloHB}

This section studies the finite sample properties of the sparse grid estimator in several Monte Carlo experiments using true random coefficient distributions of varying smoothness and dimensionality. The experiments apply the estimator to a random coefficients logit model with individual-level discrete choice data.\footnote{For a detailed description of the random coefficients multinomial logit, see \citealp[pp.~134-150]{train2009}.} The model is widely used in applied econometrics to study discrete choices of economic agents among a finite number of alternatives. 
In this model, every observation unit $n$ makes a single discrete choice among $J$ mutually exclusive alternatives (and an outside option). Observation units pick the alternative that realizes the highest utility. Let $u_{n,j} = \bm{x}_{n,j}^T\bm{\beta}_n + \omega_{n,j}$ denote the utility from alternative $j$, given covariates $\bm{x}_{n,j}$ and unobserved individual-specific preferences $\bm{\beta}_n$. The random variable $\omega_{n,j}$ denotes an additive, consumer- and  choice-specific error term. Observation unit $n$ chooses alternative $j$ if $u_{n,j} > u_{n,k}$ for all $k \neq j$ (and $u_{n,0} = \omega_{n,0}$). 
Under the assumption that $\omega_{n,j}$ is i.i.d. type I extreme value across alternatives and observation units, the unconditional choice probabilities, $P_{n,j}(\bm{x})$, are of the form 
\begin{equation}\label{logit}
P_{n,j}(\bm{x}) = \int_{\Omega_1}\cdots\int_{\Omega_D} \frac{\exp\left(\bm{x}_{n,j}^T\bm{\beta}\right)}{1 + \sum_{j = 1}^J\exp\left(\bm{x}_{n,j}^T\bm{\beta}\right)}f\left(\bm{\beta}\right)\diff\beta_D\ldots\diff\beta_1.
\end{equation}

\begin{figure}[h]%
\captionsetup{justification=centering, font=normalsize, skip=10pt} 
\caption{True and Estimated Bivariate Joint CDF of Mixture of Two Normals for $N=10,000$.}
\centering
\subfigure[True]{\includegraphics[width=0.45\textwidth]{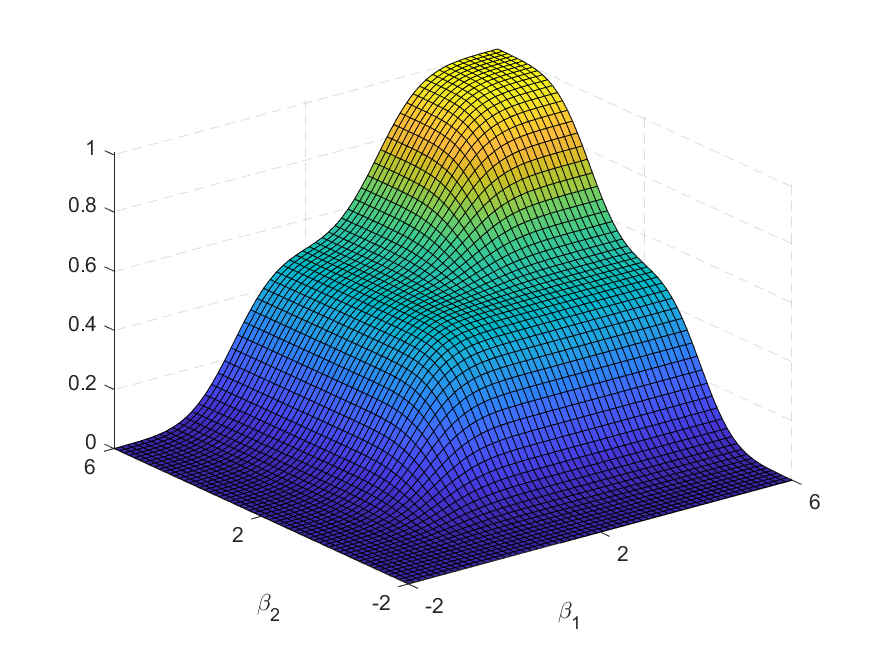}}
\subfigure[FKRB]{\includegraphics[width=0.45\textwidth]{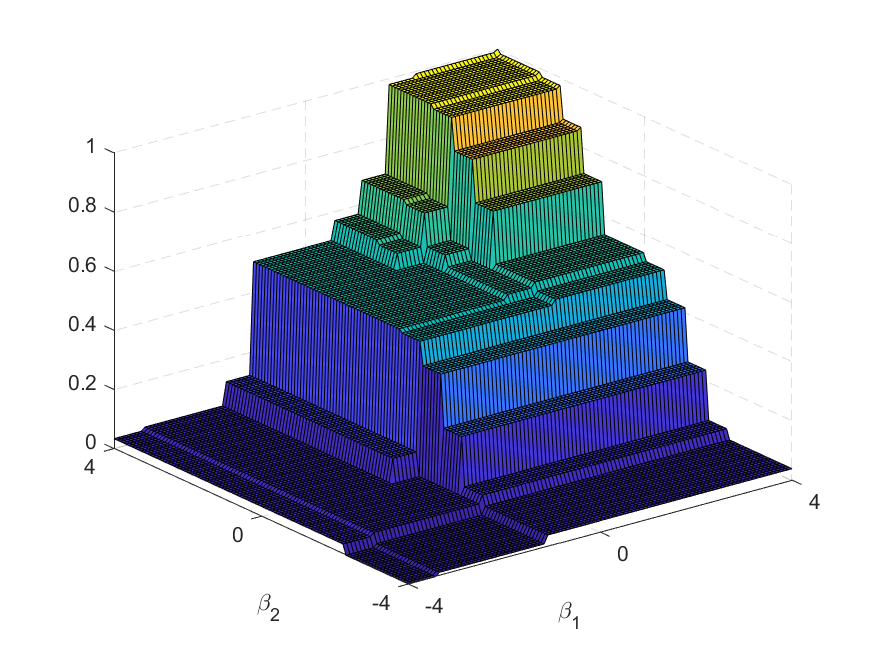}}
\subfigure[SG]{\includegraphics[width=0.45\textwidth]{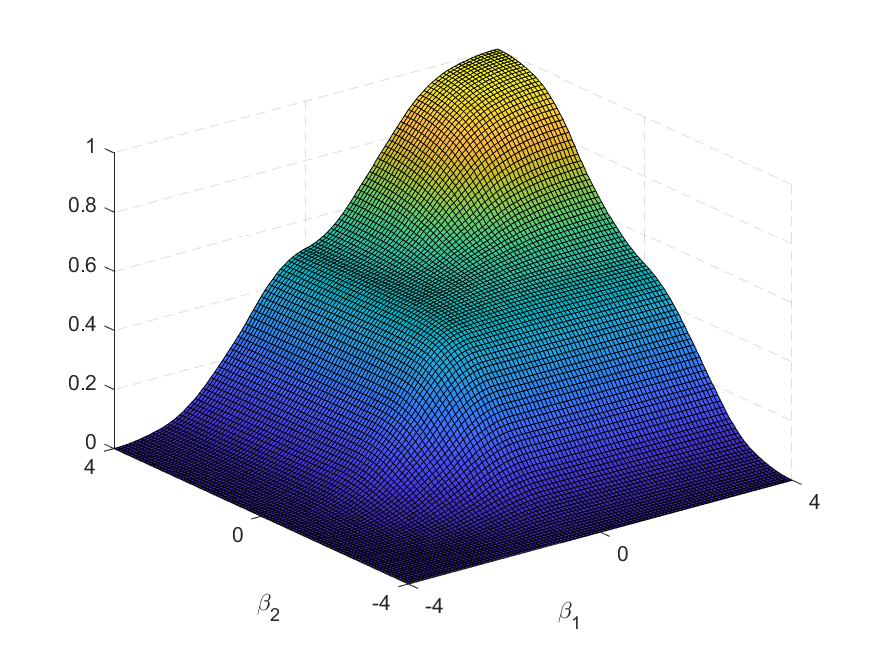}} 
\subfigure[ASG]{\includegraphics[width=0.45\textwidth]{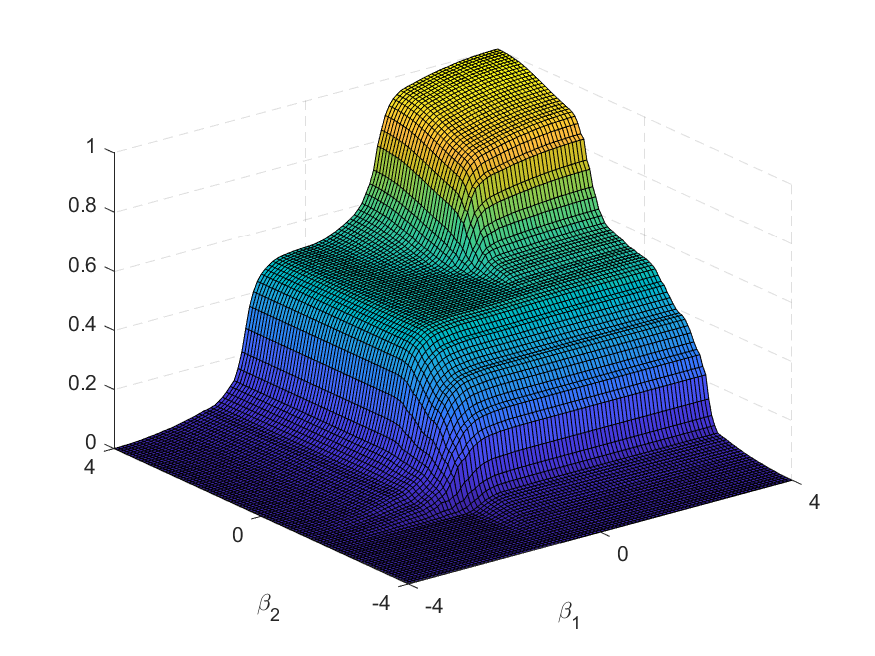}} 
\captionsetup{justification=justified,singlelinecheck=true, width=1\textwidth, font={footnotesize,stretch=0.8}} 
\caption*{\textit{Note:} Estimated bivariate joint distribution function for the mixture of two normals estimated with the FKRB estimator using 225 grid points, with the sparse grid estimator  of level $l_S=4$ (SG), and the spatially adaptive sparse grid estimator (ASG). The number of refinement steps is selected using $5$-fold cross-validation and based on the lowest out-of-sample mean squared error.}
\label{figHB:MC_True_CDFs_design1}
\end{figure}

\begin{figure}[h]%
\captionsetup{justification=centering, font=normalsize, skip=10pt} 
\caption{True and Estimated Bivariate Joint CDF of Mixture of Four Normals for $N=10,000$.}
\centering
\subfigure[True]{\includegraphics[width=0.45\textwidth]{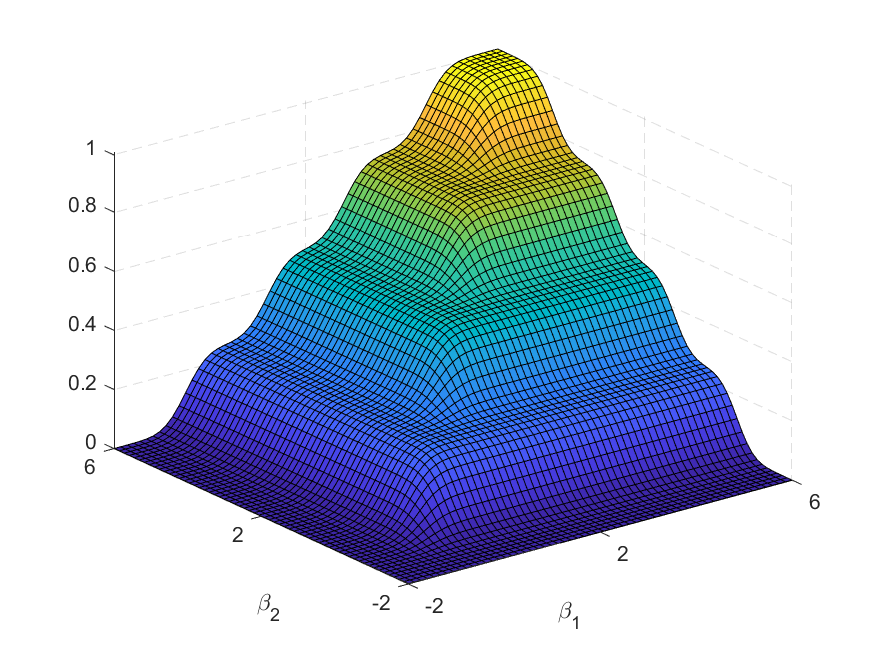}}
\subfigure[FKRB]{\includegraphics[width=0.45\textwidth]{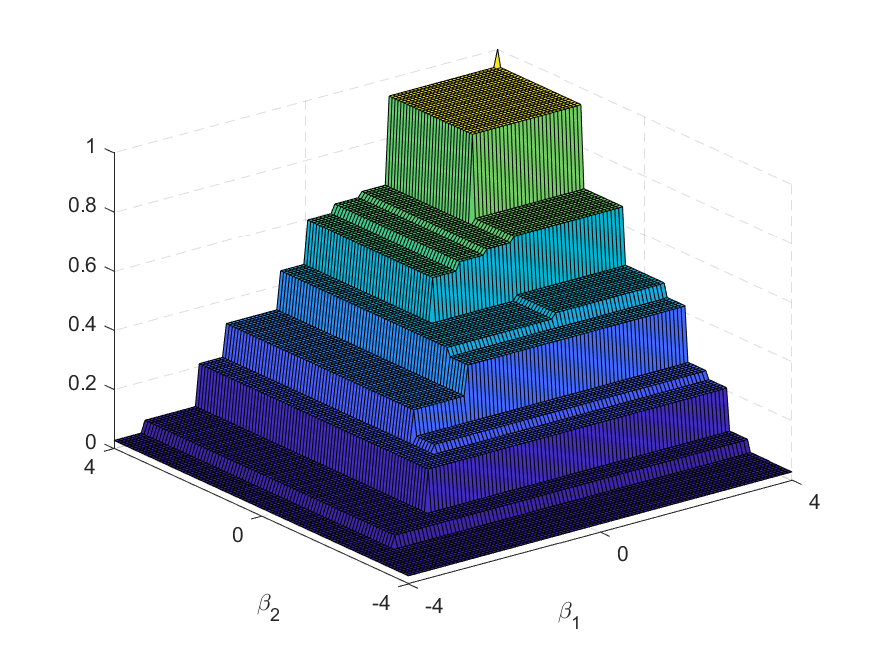}}
\subfigure[SG]{\includegraphics[width=0.45\textwidth]{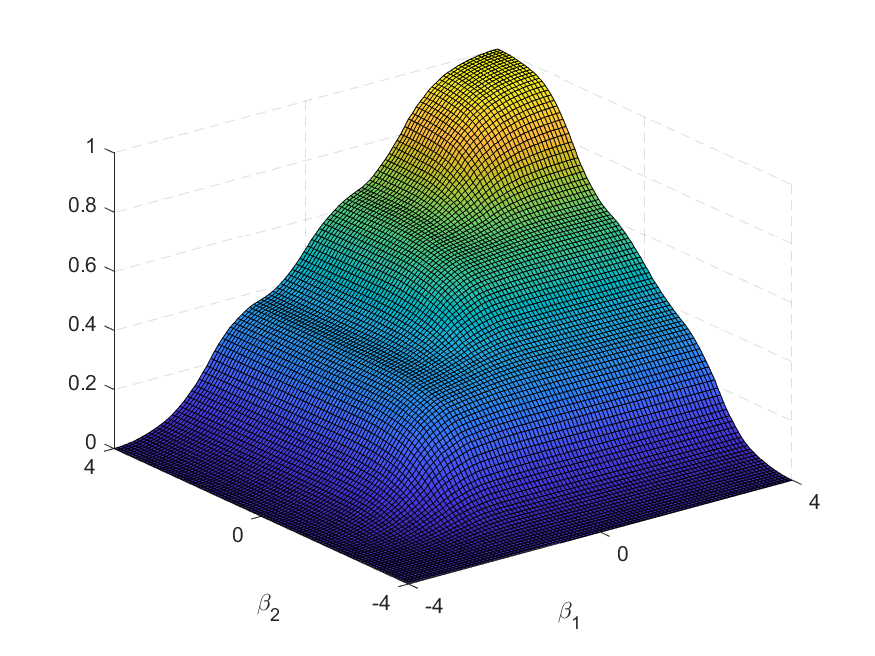}} 
\subfigure[ASG]{\includegraphics[width=0.45\textwidth]{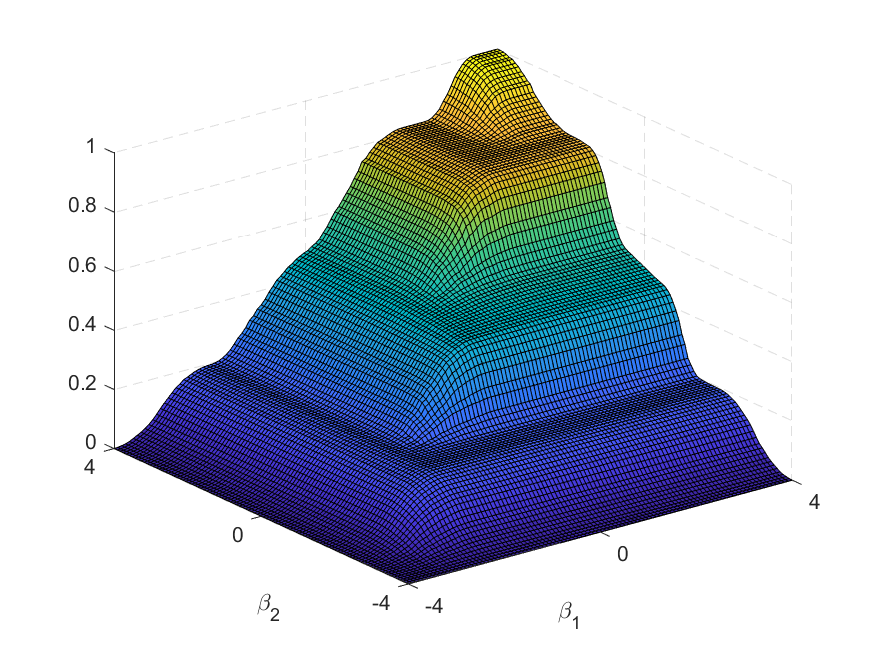}} 
\captionsetup{justification=justified,singlelinecheck=true, width=1\textwidth, font={footnotesize,stretch=0.8}} 
\caption*{\textit{Note:} Estimated bivariate joint distribution function for the mixture of four normals estimated with the FKRB estimator using 225 grid points, with the sparse grid estimator  of level $l_S=4$ (SG), and the spatially adaptive sparse grid estimator (ASG). The number of refinement steps is selected using $5$-fold cross-validation and based on the lowest out-of-sample mean squared error.}
\label{figHB:MC_True_CDFs_design2}
\end{figure}

In our experiments, the observation units choose among $J=5$ mutually exclusive alternatives and an outside option. We estimate the model for different sample sizes $N=1000, 10000$, and number of random coefficients $D=2, 4, 6$. We draw the entries of the $D$-dimensional vectors of alternative-specific characteristics, $\bm{x}_{n,j}$, independently from a $\mathcal{N}(0, 1)$ for every observation unit $n$ and alternative $j$. 
In order to study the performance of the sparse grid estimator for distributions of varying smoothness, we consider two alternative distributions for the true random coefficients distribution.
The first experiment generates the random coefficients $\bm{\beta}$ from a mixture of two multivariate normals, 
\begin{equation}
0.5 \cdot\mathcal{N}\left(\bm{\mu}^{(1)}, \bm{\Sigma}^{(1)}\right) + 0.5\cdot\mathcal{N}\left(\bm{\mu}^{(2)}, \bm{\Sigma}^{(2)}\right), \notag
\end{equation} 
where the entries of the $D$-dimensional mean vectors are $\mu_d^{(1)}=-1.5$ and $\mu_d^{(2)}=1.5$ for $d=1,\ldots,D$. The variance matrices $\bm{\Sigma}^{(1)}=\bm{\Sigma}^{(2)}$ have entries $\Sigma_{dd}^{(1)}=0.4$ on the main diagonal and $\Sigma_{dk}^{(1)}=0.1$ on the off-diagonal, i.e., for $d\neq k$.
The second experiment considers a more sophisticated and less smooth distribution. It generates the random coefficients from a mixture of four multivariate normals, 
\begin{equation}
0.25 \cdot\mathcal{N}\left(\bm{\mu}^{(1)}, \bm{\Sigma}^{(1)}\right) +  \ldots + 0.25\cdot\mathcal{N}\left(\bm{\mu}^{(4)}, \bm{\Sigma}^{(4)}\right),\notag
\end{equation} 
with $\mu_d^{(1)}=-2.5$, $\mu_d^{(2)}=-0.8$, $\mu_d^{(3)}=0.8$, and $\mu_d^{(4)}=2.5$ for $d=1,\ldots,D$. The variance matrices of the second design, $\bm{\Sigma}^{(m)}$, $m=1,\ldots,4$, are $1/4$ times the variance matrix of the first design, implying a steeper curvature.  
Figure \ref{figHB:MC_True_CDFs_design1} and Figure \ref{figHB:MC_True_CDFs_design2} display the bivariate joint distribution functions of the mixture of two normals, and the mixture of four normals, respectively. Due to the smaller variance and the higher number of mixture components, the mixture of four multivariate normals has a steeper and wigglier curvature, which, in theory, is more difficult to recover for the sparse grid estimator. Figure \ref{figHB:MC_True_PDFs} in the appendix shows the true joint probability densities of true distributions. 

For every distribution, we generate $200$ data sets for every combination of $N$ and $D$. For every data set, we estimate the random coefficients' distribution using the sparse grid and the spatially adaptive sparse grid estimator. The sparse grids are constructed on $\Omega=[-4, 4]^D$ for levels $l_S=(2,3,4)$. The support covers the true support with a coverage probability close to one (at least $0.998$). 
The hierarchical bases are constructed using piecewise $D$-linear hat functions. We simulate the integral using quasi-random number sequences. To ensure proper coverage of the true distributions' support, we let the number of simulation draws $R$ increase with the dimension of the true distribution, i.e., we use $R=D\cdot 2000$ Halton draws. We also conducted the experiments using the mexican hat function (cf. \citealp{pflueger2010}). Table \ref{tabHB:MC_Results_RMISE_mexicanHat} in the appendix presents the results which are similar to those obtained with the piecewise-linear hat function.   
For the spatially adaptive refinement, we conduct $10$ refinement steps, whereby the maximum discretization level is $5$. In every refinement step, we select the grid point (among those grid points that can be updated) with the largest contribution to the local squared error (c.f. Section \ref{secHB:adaptive}). We select the number of refinement steps using $5$-fold cross-validation, whereby the final spatially adaptive sparse grid estimator uses the refined grid that achieves the lowest out-of-sample MSE. In addition, we studied the performance of the spatially adaptive refinement when the number of refinement steps is selected based on the out-of-sample log-likelihood, and the AIC. The results are quite similar for all three criteria as indicated by the results in Table \ref{tabHB:MC_Results_RMISE_ASG} in the appendix.


As a benchmark, we estimate the random coefficients distribution using the nonparametric estimator of \citet{fox2011}. The estimator uses a fixed grid of random coefficients instead of basis functions for approximating the underlying distribution. To assure a certain comparability across the estimators in terms of the number of parameters, we use the same number of grid points, $q=3, 7, 15$, in every dimension as the full hierarchical basis has basis functions for $l=2, 3, 4$. 
We construct the $D$-dimensional grid points from the cartesian product of the one-dimensional points. This is in line with \citet{fox2011}, who recommend increasing the number of grid points exponentially with $D$. For $D=4$, we can only estimate the random coefficients distribution with $q=3, 7$ points in every dimension, and for $D=6$ with only $q=3$ grid points. Using more grid points in these setups is not possible as the number of parameters exceeds the sample size. 

We assess the estimators' approximation accuracy using the root mean integrated squared error (RMISE) from \citet{fox2011}. Denote the estimated distribution function in Monte Carlo run $m$ evaluated at $\beta_e$ by $\widehat{F}_m(\beta_e)$, and the true distribution by $F_0(\beta_e)$. The RMISE averages the squared difference between the true and estimated distribution at a fixed set of evaluation points across all Monte Carlo runs,
\begin{equation} \label{eqHB:HB_HB_RMISE}
\text{RMISE}=\sqrt{\frac{1}{200} \sum_{m=1}^{200} \left[ \frac{1}{E} \sum_{e=1}^E \left(\widehat{F}_m(\beta_e)-F_0(\beta_e)\right)^2\right]}.\notag
\end{equation}
We use a uniform grid with $E=10^D$ points spread on $\left[-4, 4\right]^D$ for the evaluation. All calculations are conducted with the statistical software R \citep{RCore}. \\

The left part in Table \ref{tabHB:MC_Results_RMISE} presents the average RMISE across the Monte Carlo replicates for the fixed grid estimator (FKRB), the sparse grid estimator (SG), and the spatially adaptive sparse grid estimator (ASG) for the mixture of two normals, while the right part the results for the mixture of four normals.
The sparse grid and the spatially adaptive sparse grid estimator achieve more accurate approximations of the true random coefficients distributions than the FKRB estimator, independent of the dimension, sample size, and the refinement level/number of fixed grid points -- even though the FKRB estimator uses a substantially greater number of parameters in higher dimensions. The difference in the approximation accuracy is particularly large for $D=6$, where the FKRB estimator cannot use more than 3 grid points in every dimension. 
The discrepancy in the approximation accuracy can be explained by the FKRB estimator's relation to the lasso (cf. \citealp{heiss2021}). Due to this relation, the estimator provides sparse solutions that lead to approximations through step functions with only a few steps. Figure \ref{figHB:MC_True_CDFs_design1} and Figure \ref{figHB:MC_True_CDFs_design2} illustrate this property for the bivariate joint mixture of two normals and bivariate joint mixture of four  normals, respectively. In contrast to the FKRB estimator, the sparse grid and the spatially adaptive sparse grid estimator provide smooth approximations due to the substantially greater number of simulation draws compared to fixed grid points used by the FKRB estimator. 
\begin{table}[H] 
\centering
\small 
\tabcolsep=0.15cm
\captionsetup{skip=0pt, justification=centering} 
\caption{Average Number of Parameters and Average RMISE over 200 Monte Carlo Replicates for Mixture of Two Normals and Mixture of Four Normals.}  \label{tabHB:MC_Results_RMISE}
\resizebox{\textwidth}{!}{\begin{tabular}{ll ccc ccc ccc ccc}  
\toprule \toprule\noalign{\smallskip}

  \\ [-2ex] 
 
& & \multicolumn{6}{c}{Mixture of two normals} & \multicolumn{6}{c}{Mixture of four normals} \\ 
\cmidrule(l{0.5pt}r{9pt}){3-8}\cmidrule(l{0.5pt}r{9pt}){9-14} 
 & & \multicolumn{3}{c}{Parameters} & \multicolumn{3}{c}{RMISE} & \multicolumn{3}{c}{Parameters} & \multicolumn{3}{c}{RMISE} \\ 
\cmidrule(l{0.5pt}r{9pt}){3-5}\cmidrule(l{0.5pt}r{9pt}){6-8}\cmidrule(l{0.5pt}r{9pt}){9-11}\cmidrule(l{0.5pt}r{9pt}){12-14} 
 $N$ & $q/l_S$ & FKRB & SG & ASG & FKRB & SG & ASG & FKRB & SG & ASG & FKRB & SG & ASG \\

 	\hline \\ [-1ex] 
 
\multicolumn{14}{l}{\textbf{Dimension} $\bm{D=2}$} \\[0.25ex] 
                                                                                                                
  1000 &  3/2 &   9 &  5 & 35.6 & 0.2067 & 0.0736 & 0.0514 &   9 &  5 & 39.1 & 0.1955 & 0.0881 & 0.0577  \\ [0ex] 
  1000 &  7/3 &  49 & 17 & 41.6 & 0.0993 & 0.0479 & 0.0536 &  49 & 17 & 48.3 & 0.1022 & 0.0473 & 0.0589  \\ [0ex] 
  1000 & 15/4 & 225 & 49 & 70.4 & 0.0912 & 0.0475 & 0.0561 & 225 & 49 & 72.9 & 0.0951 & 0.0549 & 0.0624  \\ [0ex] 
 10,000 &  3/2 &   9 &  5 & 42.5 & 0.2039 & 0.0718 & 0.0280 &   9 &  5 & 50.1 & 0.1934 & 0.0863 & 0.0435  \\ [0ex] 
 10,000 &  7/3 &  49 & 17 & 52.9 & 0.0843 & 0.0418 & 0.0290 &  49 & 17 & 63.6 & 0.0854 & 0.0434 & 0.0418  \\ [0ex] 
 10,000 & 15/4 & 225 & 49 & 76.3 & 0.0581 & 0.0313 & 0.0303 & 225 & 49 & 84.5 & 0.0648 & 0.0519 & 0.0394  \\ [0ex]

  \\ [-1ex] 
 
\multicolumn{14}{l}{\textbf{Dimension} $\bm{D=4}$} \\[0.25ex] 
                                                                                                                      
  1000 &  3/2 &   81 &   9 & 149.7 & 0.2254 & 0.0983 & 0.0583 &   81 &   9 & 153.7 & 0.2328 & 0.1201 & 0.0634  \\ [0ex] 
  1000 &  7/3 & 2401 &  49 & 213.8 & 0.1273 & 0.0620 & 0.0598 & 2401 &  49 & 226.9 & 0.1241 & 0.0850 & 0.0632  \\ [0ex] 
  1000 & 15/4 &    $-$ & 209 & 341.7 & $-$ & 0.0502 & 0.0569 &    $-$ & 209 & 341.2 & $-$ & 0.0691 & 0.0577  \\ [0ex]       
 10,000 &  3/2 &   81 &   9 & 144.1 & 0.2226 & 0.0979 & 0.0409 &   81 &   9 & 151.6 & 0.2316 & 0.1197 & 0.0536  \\ [0ex] 
 10,000 &  7/3 & 2401 &  49 & 193.8 & 0.0787 & 0.0613 & 0.0386 & 2401 &  49 & 215.2 & 0.0915 & 0.0846 & 0.0511  \\ [0ex] 
 10,000 & 15/4 &    $-$ & 209 & 341.5 & $-$ & 0.0492 & 0.0361 &    $-$ & 209 & 379.6 & $-$ & 0.0685 & 0.0480  \\ [0ex]

  \\ [-1ex] 
 
\multicolumn{14}{l}{\textbf{Dimension} $\bm{D=6}$} \\[0.25ex] 
                                                                                                                      
  1000 &  3/2 & 729 &  13 &  276.2 & 0.2156 & 0.0853 & 0.0569 & 729 &  13 &  266.4 & 0.2441 & 0.1099 & 0.0733  \\ [0ex] 
  1000 &  7/3 &   $-$ &  97 &  386.9 & $-$ & 0.0643 & 0.0547 &   $-$ &  97 &  416.8 & $-$ & 0.0909 & 0.0626  \\ [0ex]       
  1000 & 15/4 &   $-$ & 545 &  954.0 & $-$ & 0.0610 & 0.0622 &   $-$ & 545 & 1024.9 & $-$ & 0.0866 & 0.0643  \\ [0ex]       
 10,000 &  3/2 & 729 &  13 &  246.9 & 0.2138 & 0.0849 & 0.0577 & 729 &  13 &  233.7 & 0.2441 & 0.1096 & 0.0774  \\ [0ex] 
 10,000 &  7/3 &   $-$ &  97 &  207.1 & $-$ & 0.0640 & 0.0625 &   $-$ &  97 &  208.1 & $-$ & 0.0906 & 0.0888  \\ [0ex]       
 10,000 & 15/4 &   $-$ & 545 & 1055.3 & $-$ & 0.0605 & 0.0687 &   $-$ & 545 & 1099.1 & $-$ & 0.0862 & 0.0568  \\ [0ex]

 	\bottomrule\bottomrule 
 
\end{tabular}} 
\captionsetup{justification=justified,singlelinecheck=true, width=1\textwidth, font={footnotesize,stretch=0.8}, skip=2pt} 
\caption*{\textit{Note:} The table reports the total number of parameter and the RMISE for the FKRB estimator,  the sparse grid estimator (SG), and the adaptive sparse grid estimator (ASG). The adaptive sparse grid estimator performs five refinement steps, whereby the final number of refinements is determined based on the lowest out-of-sample mean squared error calculated with five-fold cross-validation. The grid point to be refined in every refinement step is selected according to its contribution to the local squared error.}
\end{table} 
Overall, the results for the sparse grid estimator presented in Table \ref{tabHB:MC_Results_RMISE} confirm the theoretical properties of the sparse hierarchical basis outlined by, e.g., \citet{bungartz2004}, as follows: (i) The estimator becomes more accurate with increasing levels -- except for $D=2$ and the mixture of four normals, where the RMISE is larger for $l_S=4$ than for $l_S=3$, which appears to be the consequence of over-fitting as indicated by the out-of-sample log-likelihood reported in Table \ref{tabHB:MC_Results_CV}. (ii) The approximation accuracy declines with an increasing number of random coefficients (except for $l_S=2$ and when going from $D=4$ to $D=6$). And (iii), the sparse grid estimator is less precise when approximating the mixture of four normals than the mixture of two normals due to the steeper and wigglier curvature of the former (except for $D=2$ and $l_S=3$) --  which is also the case for the FKRB estimator. 

The limited ability to accurately approximate non-smooth distributions is illustrated by the estimated bivariate joint CDFs in Figure \ref{figHB:MC_True_CDFs_design1} and Figure \ref{figHB:MC_True_CDFs_design2}.
The visual inspection of the estimated CDF of the mixture of two normals indicates that the sparse grid estimator is able to accurately approximate the smooth curvature of the true distribution. Inspecting the estimated mixture of four normals reveals that the estimator cannot recover the steep and wiggly shape of the true distribution. This can be explained by the limited number of basis functions with sufficiently small support in every dimension (i.e., basis functions with a high level in every dimension). The spatially adaptive sparse grid estimator, in contrast, which incrementally adds basis functions of higher levels, is able to approximate such a curvature accurately as illustrated by the estimated joint CDF.

\begin{figure}[H]
\captionsetup{justification=centering, font=normalsize, skip=10pt} 
\caption{True and Estimated Marginal CDFs of $\beta_1$ for $N=10,000$ and $l_S=4$.}
\centering
\includegraphics[width=0.9\textwidth, keepaspectratio=true]{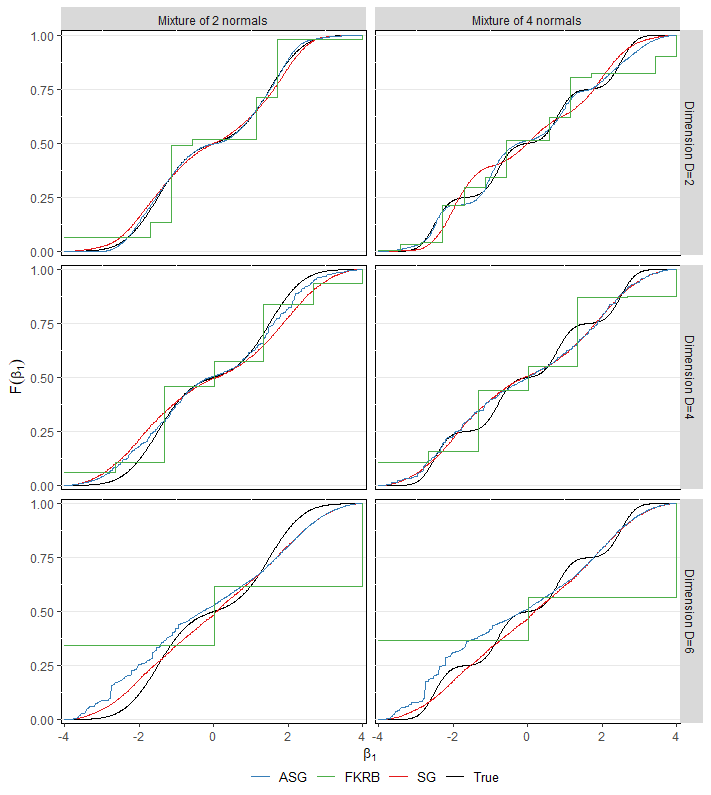}  
\captionsetup{justification=justified,singlelinecheck=true, width=1\textwidth, font={footnotesize,stretch=0.8}} 
\caption*{\textit{Note:} The figure shows the true and estimated marginal CDFs of $\beta_1$ for the mixture of two and the mixture of four normals across models with different number of random coefficients for $N=10,000$. The sparse grid estimator has level $l_S=4$, and the spatially adaptive sparse grid estimator refines the sparse grid conducting 15 refinement steps whereby the final estimator is selected based on the lowest out-of-sample MSE. The FKRB estimator estimates the two-dimensional distributions with 15 grid points, the four-dimensional distribution with 7 grid points, and the six-dimensional distribution with 3 grid points in every dimension.}
\label{figHB:MC_est_marginal_cdfs}
\end{figure}

With respect to the finite sample properties, the sparse grid estimator improves only slightly with increasing sample size for small levels $l_S=2$ and $l_S=3$, independent of the true distribution. Due to the larger support of the hierarchical basis functions with smaller levels and the imposed shape restriction following from these basis functions, the sparse hierarchical bases of levels $l_S=2$ and $l_S=3$ restrict the flexibility of the estimator to rather smooth approximations, despite an increasing sample size. 
For sparse hierarchical bases of level $l_S=4$, in contrast, the approximation accuracy of the sparse grid estimator improves stronger when increasing the sample size from $N=1000$ to $N=10,000$ due to the larger number of basis functions with small support. However, this effect declines as the number of random coefficients included in the model increases. In fact, the improvement in the RMISE is negligible for $D=6$, independent of the level and true distribution. 
The results indicate that the number of basis functions in the sparse hierarchical basis could potentially increase faster than in the classical sparse grid to obtain more accurate approximations, which would be possible with respect to the total number of parameters. 
This impression is confirmed by the estimated marginal CDFs of $\beta_1$ for the mixture of two normals and the mixture of four normals in Figure \ref{figHB:MC_est_marginal_cdfs}. Considering the mixture of two normals, the sparse grid estimator succeeds to accurately approximate the true marginal CDF for $D=2$. However, the approximation becomes less accurate as the number of random coefficients included in the model increases, indicating that there are too few hierarchical basis functions with sufficiently small support to recover the curvature of the true distribution. 
This effect is even stronger for the mixture of four normals, where the sparse grid estimator cannot recover the steep and wiggly curvature of the true distribution. While for $D=2$, the sparse grid estimator can at least recover the curvature at the boundary of the domain, for $D=4$ and $D=6$ it approximates the true marginal CDFs through a line -- though correctly located.

The results for the spatially adaptive sparse grid estimator reported in Table \ref{tabHB:MC_Results_RMISE} show that the performance of the refinement depends on the level of the sparse grid and the true distribution. First, the improvement is strongest for sparse hierarchical bases of level $l_S=2$ and declines as the level increases, independent of the shape of the true distribution. In fact, for $D=2$ and $N=1000$, the refinement leads to an improvement only for $l_S=2$, indicating that the refinement can rapidly lead to over-fitting if the dimension and sample size is small. This is also indicated by the out-of-sample MSE plotted in Figure \ref{figHB:refinement_mse_out_of_sample}, which remains more or less constant with increasing refinement steps. 
Second, the refinement is more effective for the mixture of four normals than for the mixture of two normals, as the approximation of the steep and wiggly curvature of the former requires more basis functions of smaller levels, i.e., with smaller support. 
For the mixture of two normals, the spatially adaptive refinement of the sparse grid of level $l_S=4$ on average leads to less precise estimates than the sparse grid itself. A potential explanation is that this is the consequence of an over-fitting problem as indicated by the in-sample and out-of-sample MSE plotted in Figure \ref{figHB:refinement_mse_out_of_sample} and Figure \ref{figHB:refinement_mse_in_sample}.
Figure \ref{figHB:MC_est_marginal_cdfs} illustrates the improvement of the sparse grid estimator through the spatially adaptive refinement. Considering the mixture of four normals, the estimated marginal CDF almost perfectly approximates the shape of the true distribution for $D=2$. However, the approximation accuracy declines with increasing dimensionality of the true distribution. Thus, increasing the number of refinement steps, which is close to the maximum number of 10 for $D=6$ (see Table \ref{tabHB:MC_Results_CV} in the appendix), might lead to more accurate approximations as more basis functions of smaller levels allow to recover the curvature of the mixture of four normals more precisely. 

\section{Empirical Application}\label{secHB:ApplicationHB}

In order to study the performance of the sparse hierarchical basis estimator on real data, we apply it to the setting of air pollution regulation from \citet{blundell2020}, hereafter referred to as BGL.\footnote{We gratefully thank \citet{blundell2020} for the provided \href{https://www.dropbox.com/sh/nsc1juijq91954y/AABdXoH5uFz-F8pqtEDVJDVqa?dl=0}{data and code}, and Stephan Hetzenecker, who provided a parallelized version of the code that substantially speeds up the calculations of the optimal weighting matrix and the counterfactuals.} They study the gains from dynamic enforcement of air pollution regulations using a discrete-time dynamic model of regulator and plant interactions. In this model, the regulator makes decisions regarding inspections and fines, and plants decide whether and when to invest in pollution abatement technologies. 
The quantification of the gains from dynamic enforcement of the regulation crucially depends on the estimation of plants' costs arising from compliance with the regulation. 
BGL estimate a random coefficients model to accommodate the unobserved heterogeneity of costs across plants. They estimate the five-dimensional joint distribution using the nonparametric fixed grid estimator of \citet{fox2011}. 
We apply the sparse grid estimator and the spatially adaptive sparse grid estimator to this setting and compare the estimated distribution and the results of counterfactual experiments calculated with the estimated distributions to the results of BGL.\\

The Clean Air Act and its amendments (CAAA) restrict the pollution of criteria and hazardous air pollutants through plants' in the United States to be at or below thresholds that could be achieved with the best technologies and practices. The US Environmental Protection Agency (EPA) used a dynamic enforcement regime to ensure plants' compliance with the CAAA. 
The EPA's inspections aim to uncover possible violations. Plants detected as violators will be subject to further inspections. These inspections, in turn, might uncover additional violations, leading to potential fines. Among other factors, the magnitude of fines depends on the economic benefit of the violating plant, and on the gravity of the violation. The latter is calculated from the actual or potential harm and plants' history of noncompliance. Plants can only exit violator status if they resolve all outstanding violations. The total cost of noncompliance to a violator arise from the investment cost required to resolve outstanding violations, from an increased level of oversight through the EPA, and from fines.\footnote{Costs from an increased level of oversight, i.e., from more frequent inspections, are caused by the potential shut down of production lines.}

While plants with fewer and less severe violations are designated as ``regular violators'', those with particularly severe or repeated violations can be designated as ``high priority violators'' (HPVs). The idea of this regulatory regime is to make it more costly for plants to be in HPV status: HPV undergo a higher level of oversight -- expressing itself through more frequent inspections -- are exposed to higher fines, and have to fulfill explicit deadlines to resolve all outstanding violations.  
The higher cost for HPV in comparison to regular violator are intended to encourage plants that are out of compliance to return to compliance via investments in improved processes and technologies \citep{blundell2020}.\\

BGL model the regulatory framework using a discrete-time dynamic model in which each plant plays a dynamic game with the regulator. In this game, the regulator decides whether or not to inspect a plant, and plants decide whether or not to invest in pollution abatement technologies. While the regulator wants plants to comply with the CAAA, which causes costs arising from inspections and issuing fines, plants seek to maximize their surplus.
The actions of the regulator and the investment decision of a plant within period $t$ are functions of the regulatory state $\Omega_t$, which is known to the regulator and plant at the beginning of a period. The regulatory state lists (i) a plant's EPA region, (ii) two-digit NAICS industrial sector,\footnote{The data covers the seven most polluting industrial sectors in North America defined by the North American Industry Classification System (NAICS).} (iii) expected gravity of potential violations, as measured by county non-attainment status and potential environmental damages for plants based on the county and industry, (iv) depreciated accumulated violations with a 10 percent quarterly depreciation rate, (v) regular violator or high priority violator status, and (vi) two quarterly lags of investment. While the states addressing the EPA region, the industry and the gravity of fines do not change over time, the depreciated accumulated violations, the violator status, and the  lagged investments can change from period $t$ to period $t+1$.

To incorporate plants' history of violations into the regulator's inspection policy, BGL model the probability of a plant being inspected through the regulator as a function of the regulatory state, $\mathcal{I}(\Omega)$.
The actual inspection decision $Ins$ arises stochastically. In each period, the regulator first receives an i.i.d. private information shock to the value of an inspection and then decides whether or not to inspect the plant. 
The regulator and plant then receive a compliance signal $\bm{e}_t\equiv (e_t^1, \ldots, e_t^5)$ which provides information on the presence and severity of a violation. BGL assume that $\bm{e}_t$ is a function only of the regulatory state $\Omega_t$, and the regulator's inspection policy and decisions (i.e., of the inspection probabilities $\mathcal{I}$, and the inspection decision $Ins_t$). Therefore, $\bm{e}_t$ is the predictor of compliance issues beyond the state, $Vio(\Omega, e^1)$. In addition, $e^2$ affects the fine chosen by the regulator through $Fine(\Omega, e^2)$, and $e^3$, $e^4$, and $e^5$ determines plants' transition to compliance, regular violator, and HPV status through $\tilde{\Omega}\equiv T(\Omega, e^3, e^4, e^5)$. 
Following the regulator's actions, $Ins$, $Vio$, $Fine$ and $T$, plants that are not in compliance under $\tilde{\Omega}$ make a binary decision $X\in\{0, 1\}$ of whether or not to invest in pollution abatement technologies. 

In order to avoid assumptions on the regulator's objective function,  BGL do not estimate the  regulator's utility function. Instead, they estimate plants' expectations of regulator actions using conditional choice probabilities (CCPs), and then use these probabilities to estimate plants' utility functions.
To condition on the state, they estimate the CCPs separately for plants in compliance, regular violators, and HPVs, and include indicators for two lags of investments; region; industry and gravity state dummies; and depreciated accumulated violations (for plants not in compliance).

The utility of a plant depends on the regulatory actions, the HPV status designation $HPV(\cdot)$, and the investment cost $\theta^X+\epsilon_{Xt}$,
\begin{align}
\begin{split}
U\left(\Omega,\bm{e}\right)=\ &\theta^I Ins\left(\Omega\right) + \theta^VVio\left(\Omega, e^1\right) + \theta^FFine\left(\Omega, e^2\right) + \\ &\theta^H HPV\left(T\left(\Omega, e^3, e^4, e^5\right)\right) + \theta^X + \epsilon_{Xt},
\end{split}
\end{align}
where $\epsilon_{Xt}$ is an idiosyncratic cost shock assumed to be known to the plant prior to its investment decision, and which is assumed to be i.i.d. type I extreme value. The plant chooses its investment decision in order to minimize its expected discounted sum of costs from inspections ($I$), violations ($V$), fines ($F$), designation as HPV ($H$), and investment ($X$). 

To account for unobserved heterogeneity across plants, BGL specify a random coefficients model which allows the structural parameters of the model, $\bm{\theta}\equiv(\theta^I,\theta^V,\theta^F,\theta^H,\theta^X)$, to vary across plants (but not over time).\footnote{With heterogeneous investment costs, the escalation mechanism may incentivize low-cost plants to invest in pollution abatement when they are regular violators and fines are low, while high-cost plants will wait until they become HPVs and fines are high.} 
They estimate the distribution of the random coefficients using the nonparametric fixed grid estimator of \citet{fox2011} with $10,0001$ five-dimensional grid points ${\theta}\equiv (\bm{\theta}_1, \ldots, \bm{\theta}_R)$ -- $10,000$ points from a Halton sequence and one point which corresponds to the  parameter estimates from a model with homogeneous utility parameters. They estimate the probability weights $\eta_r$, $r=1, \ldots, R$ at the grid points from the data using a GMM estimator similar to the approach of \citet{nevo2016}. The estimator minimizes the squared distance between the value of some statistic in the data, $m_k^{\text{d}}$, and the weighted sum of the statistic estimated at every grid point, $m_k(\bm{\theta}_r)$, $k=1, \ldots, K$, subject to the constraints that the weights are nonnegative and sum up to one. 
Let $G(\bm{\eta})$ denote the $K\times 1$ vector of moments with the $k$th entry being $G_k(\bm{\eta})=m^{\text{d}}_k-\sum_{r=}^R\eta_r m_k(\bm{\theta}_r)$. The GMM estimator solves the constrained optimization problem
\begin{align}\label{eqHB:GMMApplication_FKRB}
\begin{split}
&\bm{\eta} = \argmin\limits_{\bm{\eta}}G' \left(\bm{\eta}\right)\ W\ G\left(\bm{\eta}\right) \\
\text{s.t.}\quad \eta_r\geq&0, \quad  r=1, \ldots, R, \quad\text{and} \quad \sum\limits_{r = 1}^R\eta_r = 1
\end{split}
\end{align}
where $W$ is a $K\times K$ weighting matrix and $G'$ the transpose of $G$. For the estimation of the probability weights, BGL calculate three sets of moments. The first set ($5,000$ moments) represents the equilibrium share of plants being in a particular time-varying state, conditional on fixed states of region, industry, and gravity states. The second set ($4,687$ moments) multiplies the first set by the share of plants investing in this state. And the third set ($4,687$ moments) multiplies the second set by the sum of investments in the following six periods.\footnote{For the calculation of the moments, BGL solve the relevant Bellman equation and calculate $m_k(\eta_r)$ for each of the $R$ grid points. They provide a detailed description on the set of moments in their paper and more information on the calculation of the moments in the online appendix of the paper.} The second and third set of moments are intended to capture the effect of plants' investments on compliance.

Because the fixed grid approach of \citet{fox2011} treats the probability weight at every grid point as a parameter, the estimation of the random coefficients distribution involves the estimation of $10,001$ parameters.
To reduce the computational burden, and in line with the results from the Monte Carlo experiments presented in the previous section, we estimate the distribution with the sparse grid estimator of level $l_S=4$, and the spatially adaptive sparse grid estimator using the same GMM approach. The corresponding moments are of the form $G_k(\bm{\alpha})=m^{\text{d}}_k - \sum_{b=1}^B\alpha_b\sum_{r=1}^R\phi_b(\bm{\theta}_r)m_k(\bm{\theta}_r)$ for $k=1, \ldots, K$. We minimize the weighted squared sum of moments subject to the constraints $\sum_{b=1}^B\alpha_b\phi_b\left(\bm{\theta}_r\right)\geq0$, $r=1, \ldots, R$, and $\sum_{b=1}^B\alpha_b\sum_{r = 1}^R\phi_b\left(\bm{\theta}_r\right) = 1$.
For the spatial refinement of the sparse grid, we make ten refinement steps. In every step, we select the grid point with the largest contribution to the squared estimated local error. The final number of refinement steps is determined using five-fold cross-validation.\footnote{To preserve all information contained in the data, we sample all moments together that use the same first and second moments, respectively, for the calculation of further moments.} The final adaptive estimator uses the grid for which the out-of-sample mean-squared-error is lowest, which is the case after eight refinements. Figure \ref{figHB:application_first_stage_adaptive} in the appendix shows the change in the out-of-sample MSE with an increasing refinement of the sparse grid. The lowest out-of-sample MSE is achieved after 14 refinements.
To increase the efficiency of the estimator, we adopt the two-step approach of BGL. In the first step, we calculate the weighting matrix using the homogeneous parameter estimates of $\bm{\theta}$ provided by BGL. In the second step, we update $W$ using the estimated random coefficients distribution from the first step. 

\begin{figure}[h]
\captionsetup{justification=centering, font=normalsize, skip=10pt} 
\caption{Estimated Marginal CDFs for Five Utility Parameters.}
\centering
\includegraphics[width=1\textwidth]{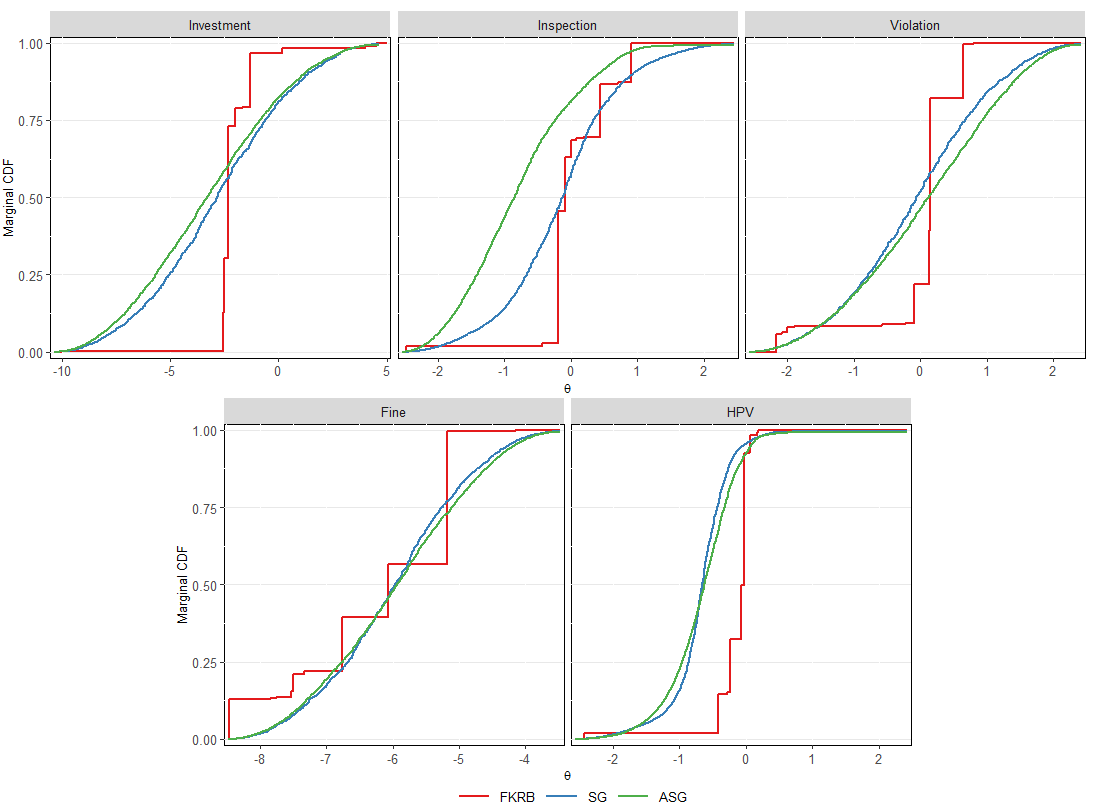}
\captionsetup{justification=justified,singlelinecheck=true, width=1\textwidth, font={footnotesize,stretch=0.8}, skip=1pt} 
\caption*{\textit{Note:} The figure shows the marginal CDFs for the five utility parameters estimated with the FKRB estimator using $10,001$ grid points, with the sparse grid estimator of level $l_S=4$ (SG), and the spatially adaptive sparse grid estimator (ASG). For the spatially adaptive sparse grid estimator, the number of refinement steps is selected using $5$-fold cross-validation and based on the lowest out-of-sample mean squared error.} 
\label{figHB:marginalCDFs_Application}
\end{figure}

Figure \ref{figHB:marginalCDFs_Application} shows the estimated marginal CDFs for the FKRB estimator, the sparse grid estimator, and the spatially adaptive sparse grid estimator.
The figure illustrates that the fixed grid estimator of \citet{fox2011} approximates the random coefficients' distribution through a step function with only a few steps -- which is a result of its sparse nature (it estimates only 14 positive weights). 
The estimated marginal CDFs of the fine, inspection and HPV utility parameters look relatively similar for all the three estimators, except that the distributions estimated with the sparse grid and spatially adaptive sparse grid estimator are much smoother. The estimated marginal CDFs of the HPV utility parameter illustrate that the sparse grid estimator can approximate a steep curvature -- though not as steep as the curvature estimated by the FKRB estimator. 
For the remaining utility parameters, the marginal CDFs estimated with the FKRB estimator and with the sparse grid and spatially adaptive sparse grid estimator deviate to a larger extent. Except for the inspection utility parameter, where the estimated marginal CDFs seem to deviate to a larger extent, the sparse grid and spatially adaptive sparse grid estimator provide similar marginal distributions. The estimated histograms plotted in Figure \ref{figHB:hist_application} in the appendix confirm the impression.

The weighted means of the estimated random coefficients' distribution reported in Table \ref{tabHB:average_utilities_Application} confirm the impression from the visual inspection of the estimated marginal CDFs. Maybe most noticeable,  plants find investments, inspections, fines, being in HPV status, and violations costly on average when the random coefficients distribution is estimated with the sparse grid estimator. This result is in line with the quasi-likelihood estimates, i.e., if plants have homogeneous utility parameters. 
For the FKRB estimator, in contrast, violations and inspections increase plants' utility on average slightly. When estimated with the adaptive sparse grid estimator, plants on average receive a positive utility from violations. 
\begin{table}[h] 
 \centering
\tabcolsep=0.4cm
\captionsetup{justification=centering, font=normalsize} 
\caption{Plants' Estimated Structural Mean Parameters.}
\label{tabHB:average_utilities_Application}
\begin{tabular}{lcccc} 
\toprule \toprule\noalign{\smallskip}  
& Quasi- & & &  \\ 
& likelihood & & &    \\ 
& estimates & FKRB & SG & ASG \\ 
\hline\\[-1ex]
                                                                                                                                    
 Negative of investment cost ($-\theta^X$) &  -2.872 & -2.051 & -2.831 & -3.232  \\[0.15ex] 
 Inspection utility ($\theta^I$) &  -0.049 &  0.047 & -0.152 & -0.797  \\[0.15ex]           
 Violation utility ($\theta^V$) & -0.077 &  0.012 & -0.062 &  0.080  \\[0.15ex]            
 Fine utility (mil. dollars, $\theta^F$)&  -5.980 & -6.211 & -5.956 & -5.920  \\[0.15ex]    
 HPV status utility ($\theta^H$)& -0.065 & -0.158 & -0.681 & -0.668  \\[0.55ex]    
 \hline
  Parameters &  5 & 10001 & 351 & 462  \\[0.15ex]          
\bottomrule\bottomrule
\end{tabular}
\captionsetup{justification=justified,singlelinecheck=true, width=0.96\textwidth, font={footnotesize,stretch=0.8}, skip=1pt} 
\caption*{\textit{Note:} The table reports the homogeneous parameter estimates (QML), and the estimated weighted mean of each random coefficient together with the total number of parameter required for the estimation of the random coefficients distribution for the FKRB estimator, the sparse gird estimator (SG), and the spatially adaptive sparse grid estimator (ASG). } 
\end{table}
Given that the estimated mean parameters are utility parameters, we can express them as fine-equivalents to compare the magnitude of the estimated means in a meaningful way. 
When the model parameters are estimated with the FKRB estimator, plants' costs for being in HPV status are equivalent to an average fine of about \$$25,516$ ($\theta^{H}/\theta^F$ multiplied by \$$1$ million), whereas the average costs for being designated as HPV are about \$$114,380$ in fine equivalents for the sparse grid estimator and about \$$112,839$ for the spatially adaptive sparse grid estimator. 
In line with that, the average costs from investments are equivalent to about \$$330,196$ in fines for the FKRB estimator, about \$$475,274$ in fines for the sparse grid estimator, and \$$545,950$  for the adaptive sparse grid estimator.
Finally, plants' average costs from inspections and violations estimated with the FKRB estimator are equivalent to fines of \$-$7,572$, and \$-$2,002$, respectively, implying that inspections and violations do not decrease utility for some plants. In contrast, inspections are equivalent to about \$$25,471$ in fines, and violations are about \$$10,405$ on average for the sparse grid estimator, and \$$134,628$ and \$-$13,514$ for the adaptive sparse grid estimator. \\

\noindent To study how the difference in the estimated distributions translates to counterfactual statistics calculated with the respective estimated random coefficients' distributions, we replicate three counterfactual experiments conducted by BGL. 
The first experiment studies how regulatory states, pollution, and investments change when the regulator fines plants in regular violator and HPV status identically for a given region, industry, and gravity state, keeping the total assessed fines the same as the baseline model for each such group. Thus, the costs of HPV status are set to zero in this experiment to fully remove dynamic enforcement. The second experiment considers the same fine structure as the first experiment but keeps the total pollution damages the same as the baseline model within each region, industry, and gravity state group. The third experiment doubles the fines for firms in HPV status compared to the baseline model, which allows to study the effect of higher escalation rates of fines.\footnote{The inspection policies in the experiments are the same as in the baseline model to assure the same state-contingent distribution of the compliance signal $\bm{e}$. Furthermore, the counterfactuals are based on surplus-optimizing plants given alternative regulatory policies and do not necessarily stem from the equilibrium of a dynamic game.}  

Table \ref{tabHB:counterfactuals1} presents the results of the experiments in terms of the long-run mean value of regulatory states, regulatory actions, investment rates, plant utility, and pollution damages for the FKRB estimator, the sparse grid estimator, and the spatially adaptive sparse grid estimator. The first column reports the long-run mean values observed in the data. The baseline columns show the outcomes calculated at the structural parameters estimated with each estimator. The estimated mean values are similar for all three estimators, and replicate the data quite well. 
\begin{table}[h] 
\centering
\small
\tabcolsep=0.2cm
\captionsetup{justification=centering} 
\caption{Counterfactual Results: Changing the Escalation Rates of Fines.}
\label{tabHB:counterfactuals1}
\begin{tabular}{lccccccc} 
\toprule \toprule\noalign{\smallskip}  
& & \multicolumn{3}{c}{\makecell{\\Baseline}} & \multicolumn{3}{c}{\makecell{Same fines for all \\ violators; fines const.}} \\ 
\cmidrule(l{0.5pt}r{9pt}){3-5}\cmidrule(l{0.5pt}r{9pt}){6-8} 
 & Data & FKRB & SG & ASG& FKRB & SG & ASG \\ 
\hline\\[-2ex]
                                                                                               
 Compliance (percent) & 95.62 & 95.11 & 95.39 & 95.35 & 66.76 & 65.47 & 80.60  \\[0.0ex]       
 Regular violator (percent) &  2.88 &  3.47 &  3.55 &  3.61 &  2.52 &  2.34 &  3.00  \\[0.0ex] 
 HPV (percent) &  1.50 &  1.42 &  1.06 &  1.04 & 30.72 & 32.19 & 16.40  \\[0.0ex]              
 Investment rate (percent) & 0.40 &  0.54 &  0.53 &  0.52 &  0.47 &  0.47 &  0.48  \\[0.0ex]   
 Inspection rate (percent) & 9.65 &  9.41 &  9.33 &  9.32 & 20.52 & 20.87 & 15.00  \\[0.0ex]   
 Fines (thousand dollars) & 0.18 &  0.32 &  0.29 &  0.28 &  0.32 &  0.29 &  0.28  \\[0.0ex]    
 Violations (percent) & 0.55 &  0.54 &  0.52 &  0.51 &  4.97 &  5.00 &  2.82  \\[0.0ex]        
 Plant utility & -- &  0.007 & -0.014 & -0.074 &  0.078 &  0.083 & -0.032  \\[0.0ex]           
 Pollution damages (mil. dollar) & 1.65 & 1.53 & 1.48 & 1.47 & 4.03 & 4.12 & 2.77  \\[0.0ex]   
\hline\\[-1ex]
& & \multicolumn{3}{c}{\makecell{Same fines for all \\violators; pollution \\damages const.}} & \multicolumn{3}{c}{\makecell{\\Fines for HPVs doubled \\ relative to baseline}} \\ 
\cmidrule(l{0.5pt}r{9pt}){3-5}\cmidrule(l{0.5pt}r{9pt}){6-8} 
 & Data & FKRB & SG & ASG& FKRB & SG & ASG \\ 
\hline\\[-2ex]
                                                                                               
 Compliance (percent) & 95.62 & 94.49 & 95.09 & 95.39 & 95.52 & 95.73 & 95.65  \\[0.0ex]       
 Regular violator (percent) &  2.88 &  2.72 &  2.27 &  2.70 &  3.47 &  3.56 &  3.62  \\[0.0ex] 
 HPV (percent) &  1.50 &  2.78 &  2.64 &  1.91 &  1.01 &  0.72 &  0.73  \\[0.0ex]              
 Investment rate (percent) & 0.40 &  0.65 &  0.70 &  0.64 &  0.55 &  0.53 &  0.52  \\[0.0ex]   
 Inspection rate (percent) & 9.65 &  9.88 &  9.80 &  9.56 &  9.28 &  9.21 &  9.21  \\[0.0ex]   
 Fines (thousand dollars) & 0.18 &  1.98 &  4.52 &  3.67 &  0.36 &  0.29 &  0.29  \\[0.0ex]    
 Violations (percent) & 0.55 &  0.74 &  0.71 &  0.63 &  0.49 &  0.46 &  0.46  \\[0.0ex]        
 Plant utility & -- &  0.001 & -0.031 & -0.089 &  0.006 & -0.015 & -0.075  \\[0.0ex]           
 Pollution damages (mil. dollar) & 1.65 & 1.53 & 1.48 & 1.47 & 1.48 & 1.44 & 1.44  \\[0.0ex]   
\bottomrule\bottomrule[-1.8ex]
\end{tabular}
\captionsetup{justification=justified,singlelinecheck=true, width=0.98\textwidth, font={footnotesize,stretch=0.8}} 
\caption*{\textit{Note:} Each statistic is the long-run equilibrium mean, weighting by the number of plants by region, industry, and gravity state in our data. Plant utility reports the average flow utility across types and states including $\epsilon$ except for Eulers constant. Column 1 presents the value of each statistic in our data. Column 2 presents the results of our model given the estimated coefficients and the existing regulatory actions and outcomes. Other columns change the state-contingent fines and the HPV cost faced by plants. Columns 3 and 4 impose the same fines for all regular and high priority violators for a given fixed state. Column 5 doubles the fines for plants in HPV status. All values are per plant/quarter.} 
\end{table}
Overall, the predicted results of the counterfactual experiments are relatively similar -- especially for the FKRB and sparse grid estimator. In the first counterfactual experiment, the share of plants in compliance predicted by the FKRB and sparse grid estimator decreases substantially from about $95\%$ to $65\%$ if fines are identical for regular violators and HPVs and the total fines are kept constant compared to the baseline model. In contrast to the FKRB estimator ($66.76\%$ in compliance and $30.72\%$ in HPV status) and the sparse grid estimator ($65.47\%$ in compliance and $32.19\%$ in HPV status), the drop in the share of plants in compliance predicted with the spatially adaptive sparse grid estimator is less strong ($80.60\%$ plants are in compliance and $16.40\%$ are designated as HPVs). 
In line with the higher share of plants in non-compliance, the predicted total pollution damages increase from $\$1.53$ mil. per plant/quarter to $\$4.03$ mil per plant quarter for the FKRB estimator and from $\$1.48$ mil. per plant/quarter to $\$4.12$ mil per plant/quarter for the sparse grid estimator. For the spatially adaptive sparse grid estimator, the total pollution damages are predicted to increase less strongly from $\$1.47$ mil. per plant/quarter to $\$2.77$ mil. per plant/quarter.  

The results of the second counterfactual experiment deviate only slightly from each other when estimated with the three estimators, except for the total fines. If the fines are the same for regular violators and high priority violators and total pollution damages are kept constant compared to the baseline model, the total amount of fines increases from $\$320$ per plant/quarter to $\$1,980$ per plant/quarter for the FKRB estimator. For the sparse grid and spatially adaptive sparse grid, the predicted total fines increase stronger from $\$290$ (SG) and $\$280$ (ASG) per plant/quarter to $\$4,520$ and $\$3,670$ per plant/quarter in comparison to the baseline model, respectively. 

For the third counterfactual experiment, the difference between mean values predicted by the FKRB, the sparse grid, and the spatially adaptive sparse grid estimator are even smaller than in the previous experiments. When the fines are doubled for HPV in comparison to the baseline model, the FKRB estimator predicts a decrease in the share of plants in compliance from $95.11\%$ to $95.52\$$, the sparse grid estimator a decrease from $95.39\%$ to $95.73\$$, and the spatially adaptive sparse grid estimator from $95.35\%$ to $95.65\$$. Thus, the substantial increase in the fines for HPVs only leads to a slight increase in the predicted share of plants in compliance. The strongest effect of the counterfactual regulatory policy is the change in the share of plants in HPV status. All three estimators predict a decrease to a similar extent (from $1.42\$$ to $1.01\%$ for the FKRB estimator, from $1.42\%$ to $0.53\%$ for the sparse grid, and from $1.04\%$ to $0.52\%$ for the spatially adaptive sparse grid estimator). Most importantly, the increased escalation of fines does only lead to a slight decrease in the predicted total pollution damages. When predicted with the FKRB estimator, the total pollution damages decrease from $\$1.53$ million to only $\$1.48$ in response to the change in the fine scheme. The predicted change is similar for the sparse grid (from $\$1.48$ million to $\$1.44$ million) and the spatially adaptive sparse grid estimator (from $\$1.47$ million to $\$1.44$ million).

\section{Conclusion}\label{secHB:ConclusionHB}

A common approach in the nonparametric literature is to approximate functions of unknown shapes using linear combinations of basis functions. For the approximation of multi-dimensional functions, the bases are typically constructed using regular tensor product constructions of one-dimensional basis functions. Such constructions lead to an exponential increase of the number of parameters in the number of dimensions, which restricts the approach to random coefficient models with only moderately few random coefficients.
In order to circumvent this limitation, we propose to use sparse hierarchical bases for the nonparametric estimation of high-dimensional random coefficient models. The proposed estimator approximates the true distribution using a linear combination of hierarchical basis functions, whereby the multi-dimensional basis functions are constructed from the one-dimensional functions using a truncated tensor product. The underlying idea goes back to \citet{smolyak1963} and has been frequently applied in mathematics and physics for the approximation of high-dimensional functions. The truncated tensor product reduces the number of basis functions substantially in comparison to a regular tensor product -- thereby rendering the estimation of high-dimensional distributions feasible. The sparse hierarchical basis deteriorates the approximation accuracy only slightly if the underlying distribution is sufficiently smooth. For non-smooth distributions, we additionally propose a spatially adaptive refinement procedure, which incrementally adds basis functions in those areas of the true distributions' domain where it has a steeper and wigglier curvature.

We study the properties of the sparse hierarchical basis estimator in various Monte Carlo experiments. Using the nonparametric fixed grid estimator of \citet{fox2011} as a benchmark, the results show that our estimator provides more accurate approximations of the true distribution, even for models with only a few random coefficients, and especially for models with moderately many random coefficients. Moreover, the results confirm the theoretical properties of sparse hierarchical bases presented by \citet{bungartz2004}. The sparse grid estimator becomes less accurate if the true distribution has a steeper and wigglier curvature and if the number of random coefficients included into the model increases. The spatially adaptive refinement of the sparse grid works particularly well for those distributions.
Applying the estimator to a data set of plants' investments in pollution abatement technologies illustrates the advantage of the sparse hierarchical basis estimator. Even though the approach requires a substantially smaller number of parameters for the estimation of the five-dimensional random coefficients distribution, the counterfactuals predicted based on the estimated distribution deviate only slightly from those predicted by the estimator of \citet{fox2011}, which involves the estimation of $10,001$ parameters. 

A practically relevant topic with respect to the application of the estimator in applied research is a valid inference procedure. Such a procedure has to take into account that the coefficients are estimated with constrained least squares, i.e., the coefficients on the boundary of the parameter space cannot have an asymptotic normal distribution.
In addition, a promising avenue for future research is to consider different kinds of sparse grids. The Monte Carlo results show that for random coefficient models with four or six random coefficients, the number of basis functions could increase faster than the rate of the classical sparse grid. Studying different kinds pf sparse grid constructions and their theoretical properties when applied to the estimation of random coefficients' distributions would provide valuable insights.

\newpage
\renewcommand\refname{References}
\bibliography{library_mendeley}

\newpage
\begin{appendix}
	
\onehalfspacing	


\setcounter{equation}{0}
\setcounter{table}{0}
\numberwithin{equation}{section}
\numberwithin{table}{section}
\numberwithin{figure}{section}
\numberwithin{subsection}{section}
\renewcommand{\thetable}{A.\arabic{table}}
\renewcommand{\theequation}{A.\arabic{equation}}
\renewcommand{\thesubsection}{A.\arabic{subsection}}
\captionsetup[figure]{list=no}
\captionsetup[table]{list=no}
\captionsetup[subsection]{list=no}

\section[Appendix A: \\ Additional Tables and Figures]{Appendix: Additional Tables and Figures}\label{appHB:tables}

\begin{figure}[H]%
\captionsetup{justification=centering, font=normalsize, skip=6pt} 
\caption{True Bivariate Probability Density Function of Mixture of Two Normals (left) and Mixture of Four Normals (right).}
\centering
\subfigure{\includegraphics[width=0.4\textwidth]{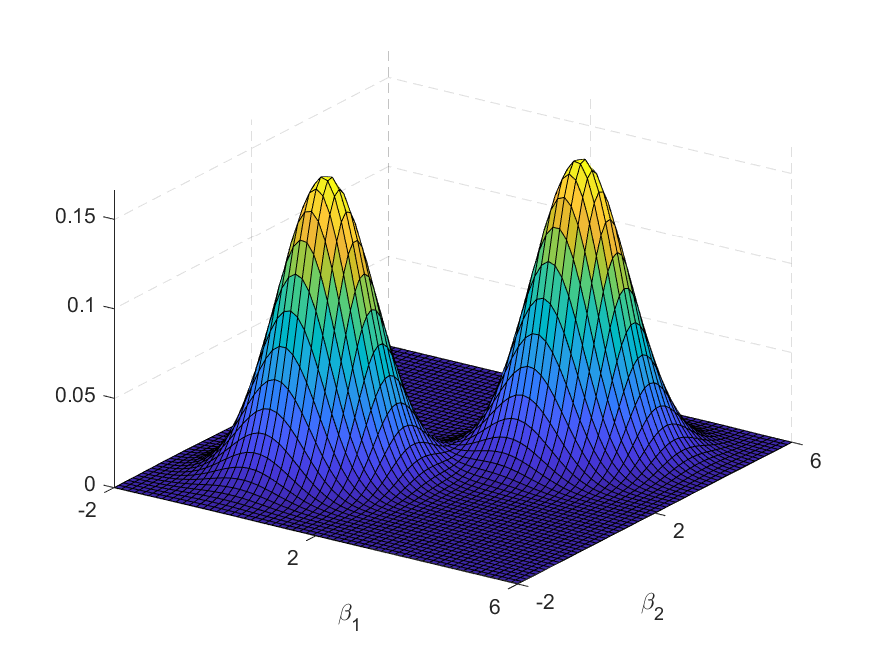}} 
\subfigure{\includegraphics[width=0.4\textwidth]{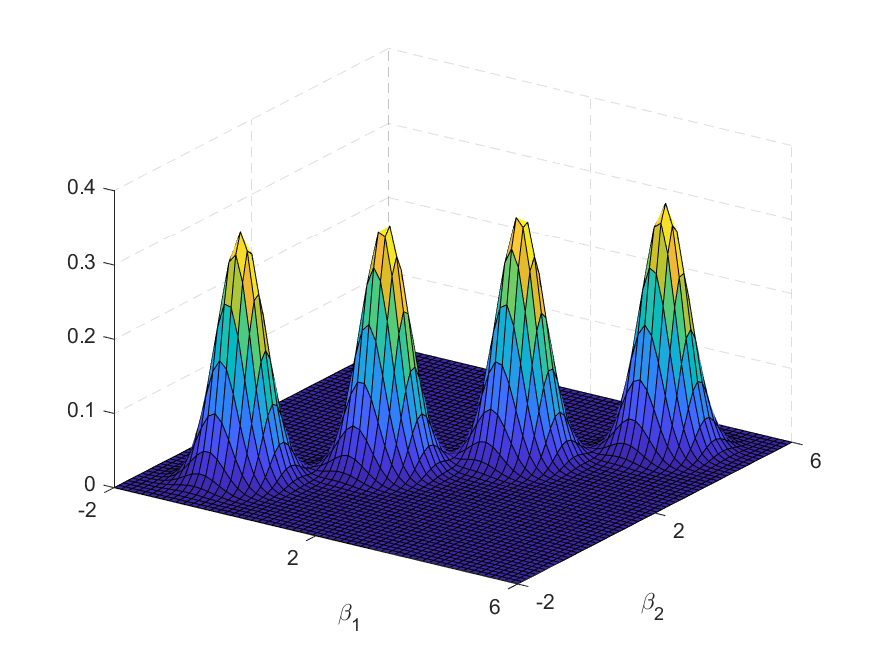}} 
\label{figHB:MC_True_PDFs}
\end{figure}

\begin{figure}[h]%
\captionsetup{justification=centering, font=normalsize, skip=6pt} 
\caption{Approximation Error of Estimated Bivariate Joint CDF of Mixture of Two Normals for $N=10,000$.}
\centering
\subfigure[FKRB]{\includegraphics[width=0.32\textwidth]{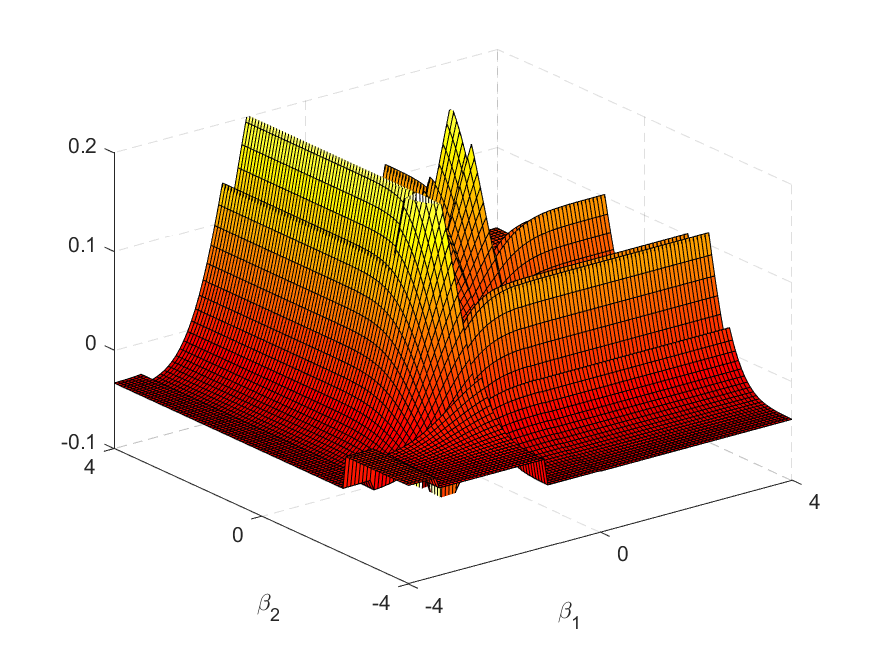}}
\subfigure[SG]{\includegraphics[width=0.32\textwidth]{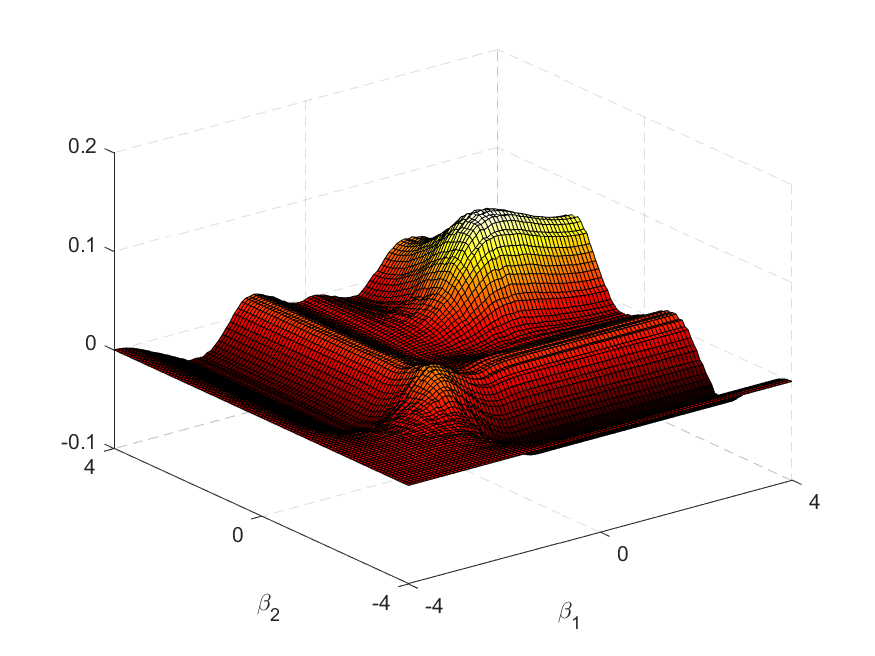}}
\subfigure[ASG]{\includegraphics[width=0.32\textwidth]{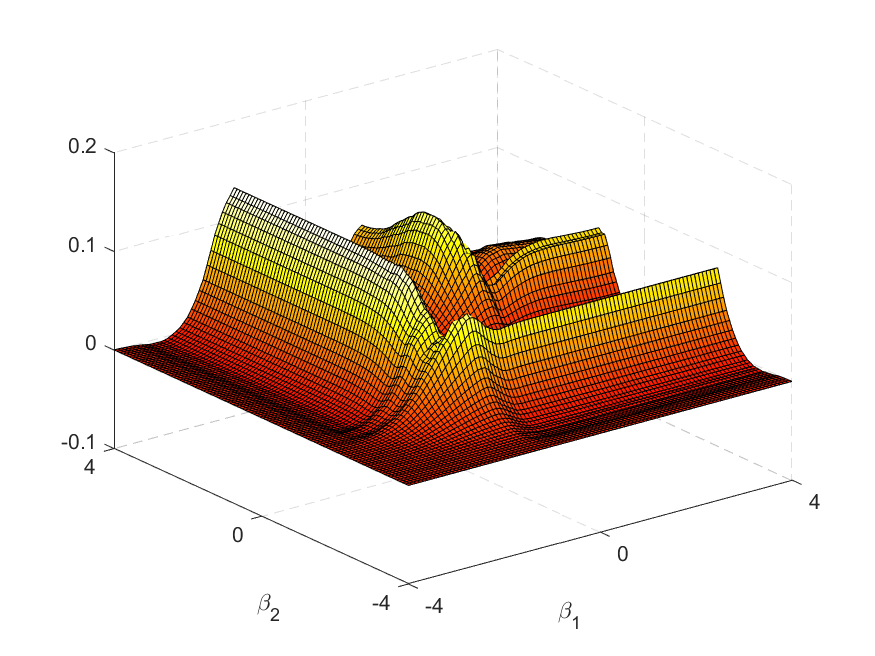}} 	
\captionsetup{justification=justified, font={footnotesize,stretch=0.8}} 
\caption*{\textit{Note:} Approximation error of estimated bivariate distribution functions of mixture of two normals for $N=10,000$, and estimated with the FKRB estimator with 225 grid points, with the sparse grid estimator with level $l_S=4$, and the spatially adaptive sparse grid estimator. The number of refinement steps is selected using $5$-fold cross-validation and based on the lowest out-of-sample mean squared error.}
\label{figHB:MC_Error_CDFs_design1}
\end{figure}

\begin{figure}[h]%
\captionsetup{justification=centering, font=normalsize, skip=6pt} 
\caption{Approximation Error of Estimated Bivariate joint CDF of Mixture of Four Normals for $N=10,000$.}
\centering
\subfigure[FKRB]{\includegraphics[width=0.32\textwidth]{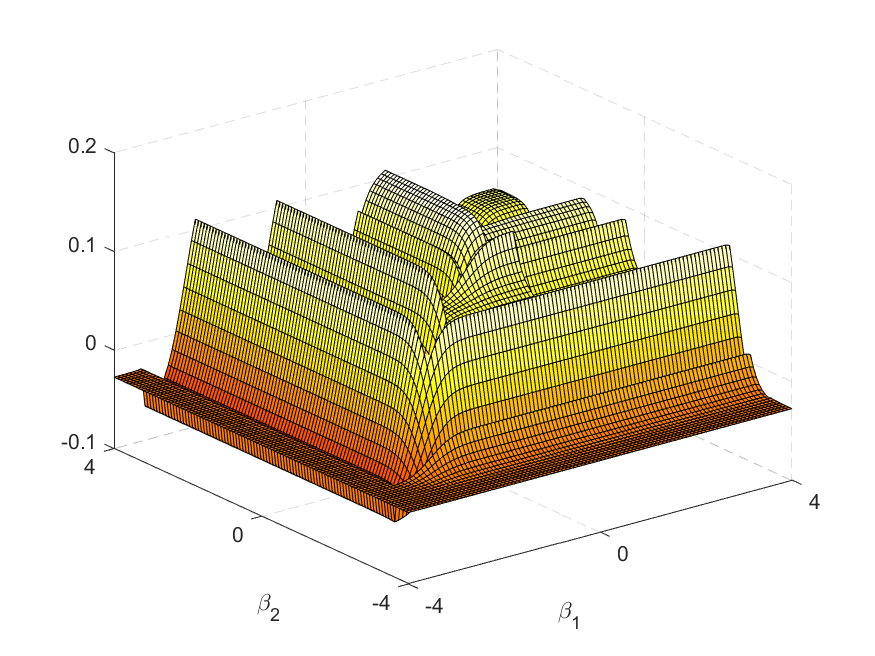}}
\subfigure[SG]{\includegraphics[width=0.32\textwidth]{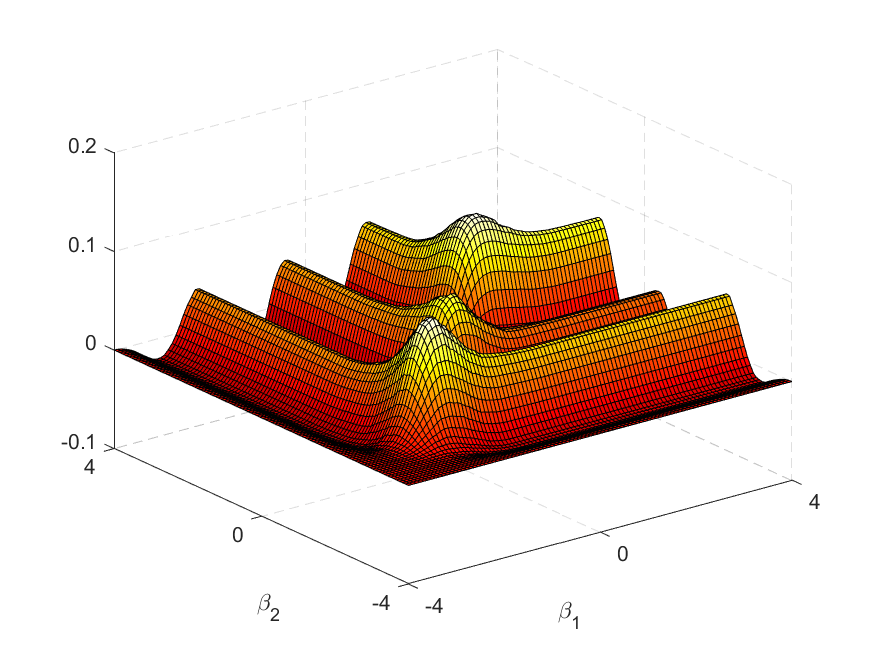}}
\subfigure[ASG]{\includegraphics[width=0.32\textwidth]{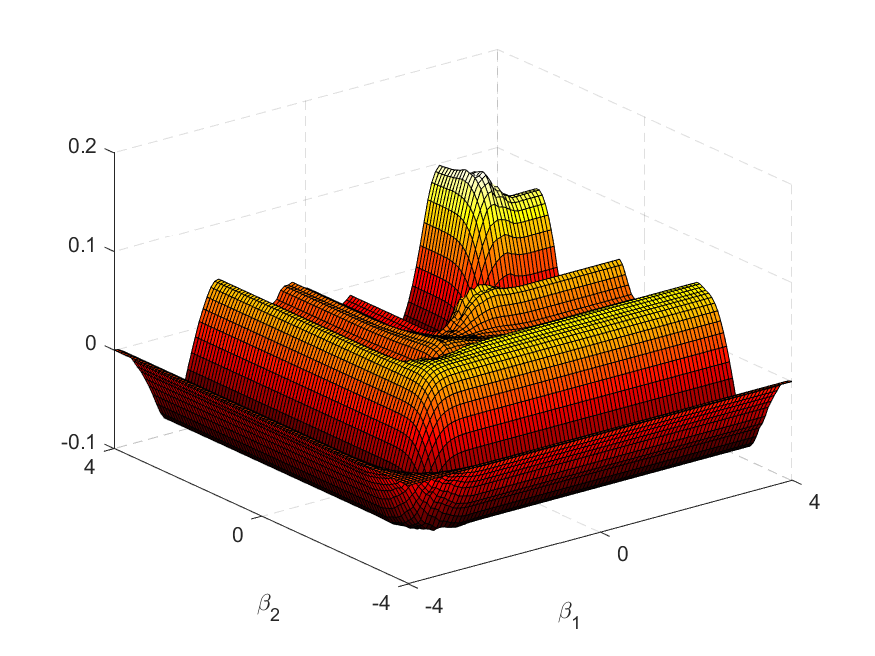}} 	
\captionsetup{justification=justified, font={footnotesize,stretch=0.8}} 
\caption*{\textit{Note:} Approximation error of estimated bivariate distribution functions of mixture of four normals for $N=10,000$, and estimated with the FKRB estimator with 225 grid points, with the sparse grid estimator with level $l_S=4$, and the spatially adaptive sparse grid estimator. The number of refinement steps is selected using $5$-fold cross-validation and based on the lowest out-of-sample mean squared error.}
\label{figHB:MC_Error_CDFs_design2}
\end{figure}

\newgeometry{left=1.5cm,right=1.5cm,top=2cm,bottom=2cm}

\begin{landscape}

\begin{table}[H] 
 \centering
\small 
\tabcolsep=0.2cm
\vspace{-0.15cm} 
\captionsetup{skip=0pt, justification=centering, font=normalsize}
\caption{Average Number of Parameters, Refinement Steps and RMISE across 200 Monte Carlo Replicates for Spatially Aadaptive Sparse Grid Estimator with Different Selection Criteria for the Number of Refinement Steps.}  
\resizebox{1.3\textwidth}{!}{ 
\begin{tabular}{ll ccc ccc ccc ccc ccc ccc}  
\toprule \toprule\noalign{\smallskip}

  \\ [-3.5ex] 
 
& & \multicolumn{9}{c}{$M = 2$} & \multicolumn{9}{c}{$M = 4$} \\ 
\cmidrule(l{0.5pt}r{9pt}){3-11}\cmidrule(l{0.5pt}r{9pt}){12-20} 
 & & \multicolumn{3}{c}{Refinements} & \multicolumn{3}{c}{Parameter} & \multicolumn{3}{c}{RMISE} & \multicolumn{3}{c}{Refinements} & \multicolumn{3}{c}{Parameter} & \multicolumn{3}{c}{RMISE} \\ 
\cmidrule(l{0.5pt}r{9pt}){3-5}\cmidrule(l{0.5pt}r{9pt}){6-8}\cmidrule(l{0.5pt}r{9pt}){9-11}\cmidrule(l{0.5pt}r{9pt}){12-14}\cmidrule(l{0.5pt}r{9pt}){15-17}\cmidrule(l{0.5pt}r{9pt}){18-20} 
 $N$ & $l_S$ & MSE & LL & AIC & MSE & LL & AIC & MSE & LL & AIC & MSE & LL & AIC & MSE & LL & AIC & MSE & LL & AIC \\

 	\hline \\ [-2ex] 
 
\multicolumn{20}{l}{\textbf{Dimension} $\bm{D=2}$} \\[0.25ex] 
                                                                                                                                                            
  1000 & 2 & 6.8 & 5.9 & 3.9 & 35.6 & 30.3 & 19.1 & 0.0514 & 0.0500 & 0.0480 & 7.3 & 6.2 & 4.0 & 39.1 & 31.9 & 19.7 & 0.0577 & 0.0568 & 0.0531  \\ [-0.1ex] 
  1000 & 3 & 4.6 & 1.9 & 0.1 & 41.6 & 26.3 & 17.3 & 0.0536 & 0.0495 & 0.0484 & 5.6 & 2.3 & 0.1 & 48.3 & 28.7 & 17.4 & 0.0589 & 0.0547 & 0.0479  \\ [-0.1ex] 
  1000 & 4 & 4.7 & 2.7 & 0.0 & 70.4 & 61.0 & 49.0 & 0.0561 & 0.0527 & 0.0475 & 5.2 & 2.6 & 0.0 & 72.9 & 60.6 & 49.0 & 0.0624 & 0.0580 & 0.0549  \\ [-0.1ex] 
 10,000 & 2 & 8.0 & 4.4 & 4.2 & 42.5 & 21.0 & 20.1 & 0.0280 & 0.0342 & 0.0351 & 9.3 & 4.6 & 4.2 & 50.1 & 22.3 & 20.4 & 0.0435 & 0.0492 & 0.0485  \\ [-0.1ex] 
 10,000 & 3 & 6.8 & 0.3 & 0.3 & 52.9 & 18.3 & 18.2 & 0.0290 & 0.0399 & 0.0401 & 9.0 & 0.9 & 0.6 & 63.6 & 20.7 & 19.2 & 0.0418 & 0.0468 & 0.0457  \\ [-0.1ex] 
 10,000 & 4 & 6.3 & 0.5 & 0.0 & 76.3 & 50.9 & 49.1 & 0.0303 & 0.0294 & 0.0312 & 8.3 & 0.4 & 0.0 & 84.5 & 50.6 & 49.1 & 0.0394 & 0.0510 & 0.0518  \\ [-0.1ex]

  \\ [-2ex] 
 
\multicolumn{20}{l}{\textbf{Dimension} $\bm{D=4}$} \\[0.25ex] 
                                                                                                                                                                  
  1000 & 2 & 9.1 & 9.1 & 6.9 & 149.7 & 149.4 &  99.8 & 0.0583 & 0.0586 & 0.0537 & 9.3 & 9.2 & 7.3 & 153.7 & 152.5 & 105.3 & 0.0634 & 0.0635 & 0.0624  \\ [-0.1ex] 
  1000 & 3 & 8.4 & 8.5 & 3.7 & 213.8 & 216.1 & 104.1 & 0.0598 & 0.0607 & 0.0530 & 8.5 & 8.7 & 3.8 & 226.9 & 231.3 & 105.1 & 0.0632 & 0.0635 & 0.0636  \\ [-0.1ex] 
  1000 & 4 & 6.5 & 6.5 & 1.3 & 341.7 & 339.3 & 222.3 & 0.0569 & 0.0582 & 0.0482 & 6.5 & 6.8 & 1.8 & 341.2 & 350.0 & 229.9 & 0.0577 & 0.0583 & 0.0545  \\ [-0.1ex] 
 10,000 & 2 & 9.3 & 8.9 & 8.6 & 144.1 & 133.6 & 126.2 & 0.0409 & 0.0423 & 0.0429 & 9.5 & 9.5 & 8.7 & 151.6 & 151.2 & 129.4 & 0.0536 & 0.0536 & 0.0529  \\ [-0.1ex] 
 10,000 & 3 & 8.3 & 7.3 & 6.6 & 193.8 & 167.5 & 151.3 & 0.0386 & 0.0409 & 0.0420 & 9.1 & 8.9 & 6.9 & 215.2 & 210.2 & 159.8 & 0.0511 & 0.0511 & 0.0501  \\ [-0.1ex] 
 10,000 & 4 & 6.8 & 5.2 & 3.6 & 341.5 & 303.3 & 265.0 & 0.0361 & 0.0381 & 0.0405 & 8.4 & 7.6 & 3.6 & 379.6 & 360.8 & 265.8 & 0.0480 & 0.0480 & 0.0472  \\ [-0.1ex]

  \\ [-2ex] 
 
\multicolumn{20}{l}{\textbf{Dimension} $\bm{D=6}$} \\[0.25ex] 
                                                                                                                                                                          
  1000 & 2 &  9.7 &  9.7 & 7.0 &  276.2 &  280.0 & 154.9 & 0.0569 & 0.0573 & 0.0564 &  9.8 &  9.8 & 7.5 &  266.4 &  267.6 & 170.8 & 0.0733 & 0.0734 & 0.0795  \\ [-0.1ex] 
  1000 & 3 &  9.2 &  9.5 & 0.1 &  386.9 &  397.9 &  99.2 & 0.0547 & 0.0552 & 0.0642 &  9.4 &  9.6 & 1.0 &  416.8 &  421.6 & 127.1 & 0.0626 & 0.0627 & 0.0869  \\ [-0.1ex] 
  1000 & 4 &  7.9 &  8.2 & 2.2 &  954.0 &  974.6 & 613.3 & 0.0622 & 0.0642 & 0.0451 &  8.7 &  8.8 & 2.8 & 1024.9 & 1028.7 & 638.4 & 0.0643 & 0.0644 & 0.0609  \\ [-0.1ex] 
 10,000 & 2 & 10.0 & 10.0 & 9.5 &  246.9 &  247.6 & 225.2 & 0.0577 & 0.0577 & 0.0562 & 10.0 & 10.0 & 9.7 &  233.7 &  233.8 & 222.1 & 0.0774 & 0.0774 & 0.0774  \\ [-0.1ex] 
 10,000 & 3 &  9.7 &  9.8 & 0.8 &  207.1 &  208.3 & 108.7 & 0.0625 & 0.0625 & 0.0636 &  9.8 &  9.9 & 1.2 &  208.1 &  209.1 & 113.4 & 0.0888 & 0.0888 & 0.0899  \\ [-0.1ex] 
 10,000 & 4 &  9.5 &  9.4 & 3.1 & 1055.3 & 1054.3 & 657.6 & 0.0687 & 0.0688 & 0.0339 &  9.8 &  9.8 & 3.9 & 1099.1 & 1102.5 & 708.9 & 0.0568 & 0.0569 & 0.0429  \\ [-0.1ex]

 	\bottomrule\bottomrule \\[-2.5ex] 
 
\end{tabular} } 
\captionsetup{justification=justified,singlelinecheck=true, width=1.285\textwidth, font={footnotesize,stretch=0.8}} 
\caption*{\textit{Note:} The table reports the total number of parameter and the RMISE for the spatially adaptive sparse grid estimator using the out-of-sample mean squared error (MSE), the out-of-sample log-likelihood (LL), and the Akaike information criterion (AIC) for the selection of the final number of refinement steps. For every estimator five refinement steps are performed. The out-of-sample mean squared error and the out-of-sample log-likelihood are calculated with five-fold cross-validation. The grid point to be refined in every refinement step is selected according to its contribution to the local squared error.} 
\label{tabHB:MC_Results_RMISE_ASG} 
\end{table}


\newpage

\begin{table}[H] 
 \centering
\small 
\tabcolsep=0.2cm
\vspace{-0.15cm} 
\captionsetup{skip=0pt, justification=centering, font=normalsize, skip=6pt}
\caption{Average Out-of-sample Mean Squared Error and Out-of-sample Log-likelihood across 200 Monte Carlo Replicates for the Sparse Grid and Spatially Adaptive Sparse Grid Estimator with Different Selection Criteria for the Number of Refinement Steps.}  \label{tabHB:MC_Results_CV}
\resizebox{1.3\textwidth}{!}{ 
\begin{tabular}{ll cccc cccc cccc cccc}  
\toprule \toprule\noalign{\smallskip}

  \\ [-3.5ex] 
 
& & \multicolumn{8}{c}{$M = 2$} & \multicolumn{8}{c}{$M = 4$} \\ 
\cmidrule(l{0.5pt}r{9pt}){3-10}\cmidrule(l{0.5pt}r{9pt}){11-18} 
 & & \multicolumn{4}{c}{MSE} & \multicolumn{4}{c}{LL} & \multicolumn{4}{c}{MSE} & \multicolumn{4}{c}{LL} \\ 
\cmidrule(l{0.5pt}r{9pt}){3-6}\cmidrule(l{0.5pt}r{9pt}){7-10}\cmidrule(l{0.5pt}r{9pt}){11-14}\cmidrule(l{0.5pt}r{9pt}){15-18} 
 $N$ & $l_S$ & SG & MSE & LL & AIC & SG & MSE & LL & AIC & SG & MSE & LL & AIC & SG & MSE & LL & AIC  \\

 	\hline \\ [-2ex] 
 
\multicolumn{18}{l}{\textbf{Dimension} $\bm{D=2}$} \\[0.25ex] 
                                                                                                                                                                                
  1000 & 2 & 1422.7 & 1367.0 & 1367.4 & 1369.1 &  -290.7 &  -276.2 &  -276.1 &  -276.4 & 1425.3 & 1358.9 & 1359.5 & 1361.5 &  -291.8 &  -274.1 &  -274.0 &  -274.4  \\ [-0.1ex] 
  1000 & 3 & 1368.7 & 1366.6 & 1367.3 & 1368.7 &  -276.2 &  -276.2 &  -275.9 &  -276.2 & 1361.0 & 1358.2 & 1359.2 & 1361.0 &  -274.2 &  -274.1 &  -273.9 &  -274.1  \\ [-0.1ex] 
  1000 & 4 & 1369.0 & 1366.8 & 1367.3 & 1369.0 &  -276.6 &  -276.4 &  -276.3 &  -276.6 & 1360.5 & 1358.1 & 1358.8 & 1360.5 &  -274.4 &  -274.3 &  -274.2 &  -274.4  \\ [-0.1ex] 
 10,000 & 2 & 1420.9 & 1366.5 & 1366.9 & 1367.0 & -2902.1 & -2766.4 & -2763.2 & -2763.3 & 1423.0 & 1357.3 & 1358.2 & 1358.4 & -2911.0 & -2741.5 & -2738.2 & -2738.3  \\ [-0.1ex] 
 10,000 & 3 & 1367.1 & 1366.4 & 1367.0 & 1367.1 & -2762.6 & -2767.0 & -2762.5 & -2762.5 & 1359.0 & 1357.0 & 1358.5 & 1358.7 & -2738.2 & -2743.4 & -2737.9 & -2738.0  \\ [-0.1ex] 
 10,000 & 4 & 1366.8 & 1366.4 & 1366.7 & 1366.8 & -2765.9 & -2767.8 & -2765.8 & -2765.9 & 1357.7 & 1356.8 & 1357.6 & 1357.7 & -2740.9 & -2744.9 & -2740.8 & -2740.9  \\ [-0.1ex]

  \\ [-2ex] 
 
\multicolumn{18}{l}{\textbf{Dimension} $\bm{D=4}$} \\[0.25ex] 
                                                                                                                                                                                
  1000 & 2 & 1471.3 & 1351.1 & 1351.2 & 1356.7 &  -299.6 &  -266.2 &  -266.2 &  -267.9 & 1469.6 & 1333.4 & 1333.5 & 1338.4 &  -299.7 &  -262.2 &  -262.1 &  -263.6  \\ [-0.1ex] 
  1000 & 3 & 1390.3 & 1348.8 & 1348.9 & 1358.8 &  -278.0 &  -265.4 &  -265.4 &  -268.7 & 1384.4 & 1330.8 & 1331.0 & 1343.2 &  -276.7 &  -261.4 &  -261.3 &  -265.1  \\ [-0.1ex] 
  1000 & 4 & 1367.4 & 1345.6 & 1345.9 & 1350.9 &  -270.9 &  -264.6 &  -264.5 &  -266.1 & 1358.3 & 1326.2 & 1326.6 & 1332.9 &  -269.0 &  -260.0 &  -259.9 &  -261.9  \\ [-0.1ex] 
 10,000 & 2 & 1468.9 & 1344.7 & 1344.7 & 1344.8 & -2986.8 & -2645.8 & -2645.6 & -2645.7 & 1466.8 & 1326.6 & 1326.6 & 1326.9 & -2991.1 & -2601.8 & -2601.7 & -2602.4  \\ [-0.1ex] 
 10,000 & 3 & 1388.9 & 1344.2 & 1344.3 & 1344.5 & -2776.7 & -2645.7 & -2645.3 & -2645.7 & 1382.6 & 1324.9 & 1324.9 & 1325.4 & -2762.6 & -2597.7 & -2597.6 & -2598.9  \\ [-0.1ex] 
 10,000 & 4 & 1364.8 & 1343.5 & 1343.7 & 1344.0 & -2703.8 & -2645.4 & -2644.8 & -2645.6 & 1355.1 & 1323.2 & 1323.3 & 1324.0 & -2682.5 & -2595.2 & -2594.9 & -2596.5  \\ [-0.1ex]

  \\ [-2ex] 
 
\multicolumn{18}{l}{\textbf{Dimension} $\bm{D=6}$} \\[0.25ex] 
                                                                                                                                                                                
  1000 & 2 & 1497.8 & 1384.8 & 1384.8 & 1404.6 &  -305.1 &  -274.5 &  -274.5 &  -280.0 & 1492.5 & 1369.0 & 1369.0 & 1385.0 &  -303.9 &  -270.6 &  -270.6 &  -275.0  \\ [-0.1ex] 
  1000 & 3 & 1419.0 & 1366.7 & 1366.8 & 1418.2 &  -284.1 &  -269.5 &  -269.5 &  -283.9 & 1409.9 & 1345.8 & 1345.9 & 1398.5 &  -281.8 &  -264.2 &  -264.2 &  -278.7  \\ [-0.1ex] 
  1000 & 4 & 1411.6 & 1356.6 & 1356.7 & 1375.3 &  -281.5 &  -266.7 &  -266.6 &  -272.0 & 1401.7 & 1335.2 & 1335.3 & 1349.0 &  -279.0 &  -261.3 &  -261.3 &  -265.3  \\ [-0.1ex] 
 10,000 & 2 & 1493.8 & 1390.6 & 1390.6 & 1390.9 & -3043.6 & -2760.5 & -2760.5 & -2761.7 & 1488.2 & 1379.1 & 1379.1 & 1379.4 & -3033.8 & -2733.3 & -2733.3 & -2734.1  \\ [-0.1ex] 
 10,000 & 3 & 1416.2 & 1413.1 & 1413.1 & 1415.1 & -2834.2 & -2823.9 & -2823.8 & -2830.7 & 1407.1 & 1403.5 & 1403.5 & 1405.5 & -2811.5 & -2799.9 & -2799.9 & -2806.8  \\ [-0.1ex] 
 10,000 & 4 & 1406.4 & 1356.7 & 1356.7 & 1359.7 & -2802.3 & -2667.6 & -2667.5 & -2677.7 & 1396.4 & 1334.9 & 1334.9 & 1340.1 & -2777.8 & -2612.6 & -2612.6 & -2628.2  \\ [-0.1ex] 
 
 \bottomrule\bottomrule \\[-2.5ex] 
 
\end{tabular} } 
\captionsetup{justification=justified,singlelinecheck=true, width=1.285\textwidth, font={footnotesize,stretch=0.8}} 
\caption*{\textit{Note:} The table reports the average out-of-sample mean squared error (MSE) and the average out-of-sample log-likelihood (LL) for the spatially adaptive sparse grid estimator using the out-of-sample mean squared error (MSE), the out-of-sample log-likelihood (LL), and the Akaike information criterion (AIC) for the selection of the final number of refinement steps. For every estimator five refinement steps are performed. The out-of-sample mean squared error and the out-of-sample log-likelihood are calculated with five-fold cross-validation. The grid point to be refined in every refinement step is selected according to its contribution to the local squared error.} 
\end{table}

\end{landscape}	


\restoregeometry

\begin{figure}[H]
\captionsetup{skip=0pt, justification=centering, font=normalsize, skip=6pt}
\caption{True and Estimated Marginal CDFs of $\beta_1$ for Mixture of Two Normals and Mixture of Four Normals for Spatially Adaptive Sparse Grid Estimator with Different Selection Criteria for $N=10,000$.} 
\centering
\includegraphics[width=1\textwidth, keepaspectratio=true]{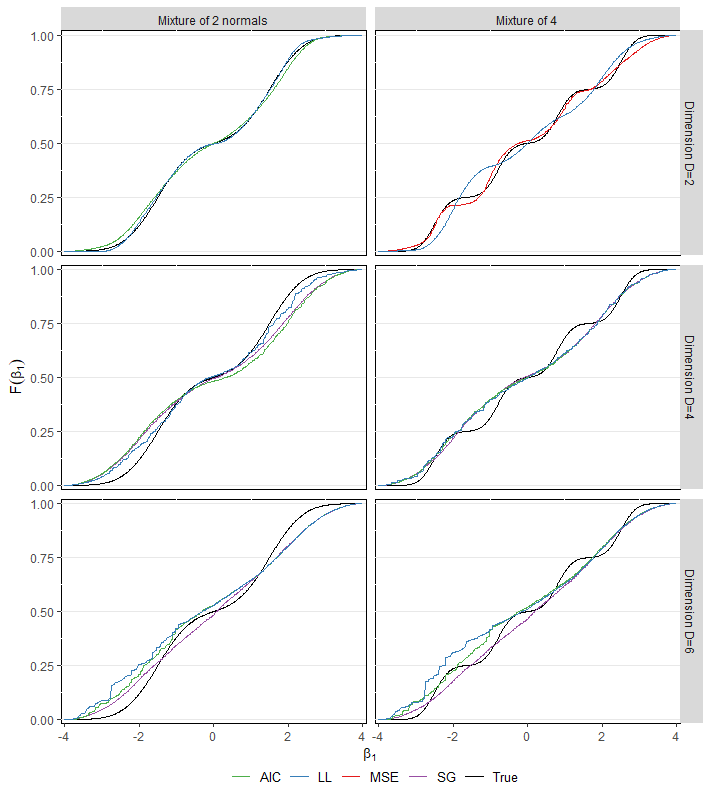}  
\captionsetup{justification=justified,singlelinecheck=true, width=1\textwidth, font={footnotesize,stretch=0.8}} 
\caption*{\textit{Note:} The figure shows the true and estimated marginal CDFs of $\beta_1$ for the mixture of two and the mixture of four normals across models with different number of random coefficients for $N=10,000$. The spatially adaptive sparse grid estimator refines a sparse grid of level $l_S=4$ conducting 15 refinement steps and using the out-of-sample MSE, the out-of-sample log-likelihood (LL), and the Akaike information criterion (AIC) to select the number of refinement steps.}\label{figHB:MC_est_marginal_cdfs_adaptive}
\end{figure}

\newpage

\begin{figure}[H]
\captionsetup{skip=0pt, justification=centering, font=normalsize, skip=6pt}
\caption{Average Out-of-sample Mean Squared Error of Spatially Adaptive Refinement across 200 Monte Carlo Replicates.} 
\centering
\includegraphics[width=1\textwidth, keepaspectratio=true]{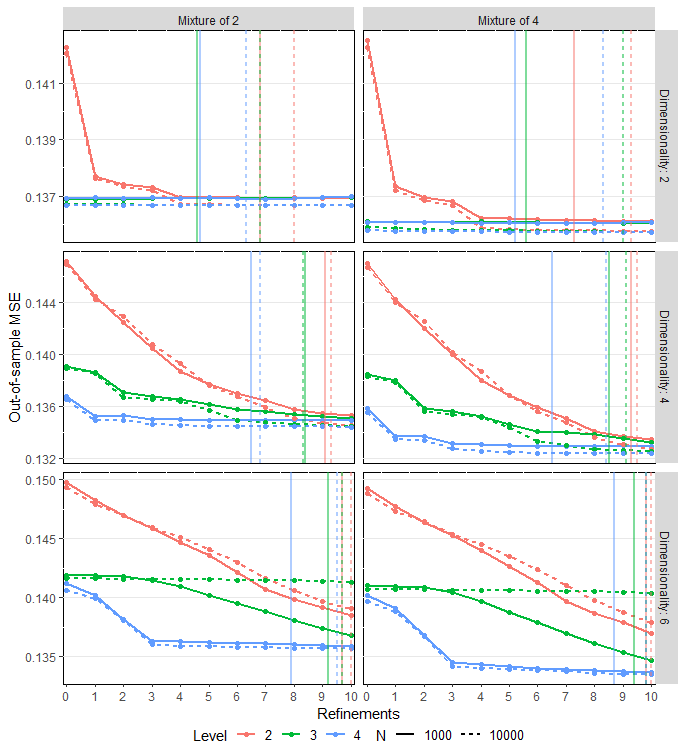}  
\captionsetup{justification=justified,singlelinecheck=true, width=1\textwidth, font={footnotesize,stretch=0.8}} 
\caption*{\textit{Note:} The figure shows the average out-of-sample mean squared error (MSE) of the spatially adaptive refinement of sparse grids of different levels across refinement steps that is calculated via $5$-fold cross-validation. The vertical lines report the average number of refinement steps that are selected based on the lowest out-of-sample MSE. The solid lines report the results for $N=1000$, the dashed lines the results for $N=10,000$.}\label{figHB:refinement_mse_out_of_sample}
\end{figure}

\newpage

\begin{figure}[H]
\captionsetup{skip=0pt, justification=centering, font=normalsize, skip=6pt}
\caption{Average In-sample Mean Squared Error of Spatially Adaptive Refinement across 200 Monte Carlo Replicates.}
\centering
\includegraphics[width=1\textwidth, keepaspectratio=true]{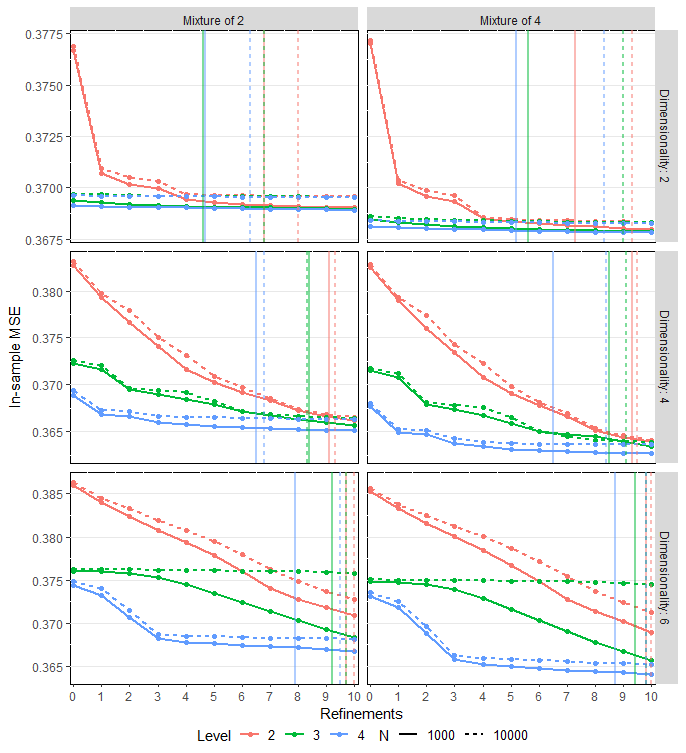}  
\captionsetup{justification=justified,singlelinecheck=true, width=1\textwidth, font={footnotesize,stretch=0.8}} 
\caption*{\textit{Note:} The figure shows the average in-sample mean squared error (MSE) of the spatially adaptive refinement of sparse grids of different levels across refinement steps that is calculated via $5$-fold cross-validation. The vertical lines report the average number of refinement steps that are selected based on the lowest out-of-sample MSE. The solid lines report the results for $N=1000$, the dashed lines the results for $N=10,000$.}\label{figHB:refinement_mse_in_sample}
\end{figure}

\newpage

\begin{figure}[H]
\captionsetup{skip=0pt, justification=centering, font=normalsize, skip=6pt}
\caption{Average Akaike Information Criterion of Spatially Adaptive Refinement across 200 Monte Carlo Replicates.}
\centering
\includegraphics[width=1\textwidth, keepaspectratio=true]{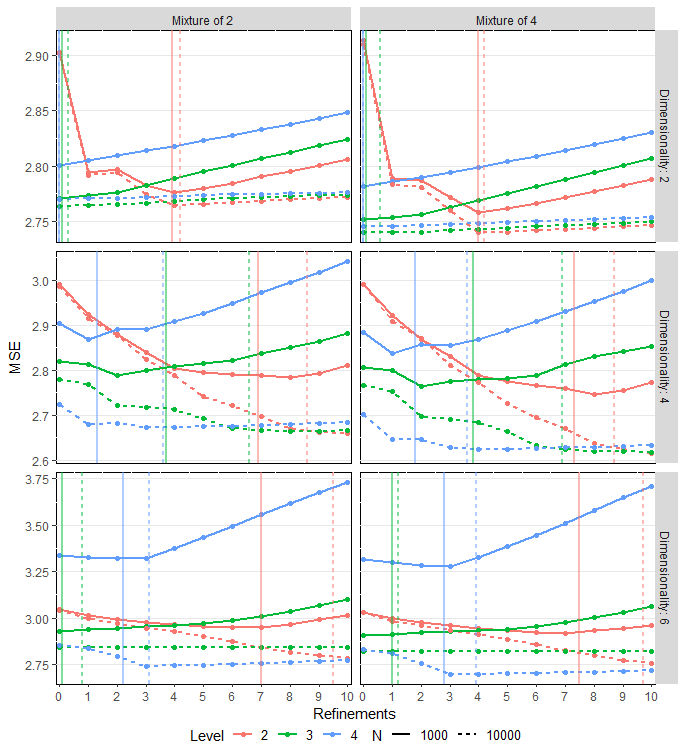}  
\captionsetup{justification=justified,singlelinecheck=true, width=1\textwidth, font={footnotesize,stretch=0.8}} 
\caption*{\textit{Note:} The figure shows the average Akaike information criterion (AIC) of the spatially adaptive refinement of sparse grids of different levels across refinement steps that is calculated via $5$-fold cross-validation. The vertical lines report the average number of refinement steps that are selected based on the lowest AIC. The solid lines report the results for $N=1000$, the dashed lines the results for $N=10,000$.}\label{figHB:refinement_mse}
\end{figure}

\newpage

\begin{table}[H] 
\label{tabHB:MC_Results_RMISE_mexicanHat}
 \centering
\small 
\tabcolsep=0.15cm
\captionsetup{skip=0pt, justification=centering} 
\caption{Total number of parameter and average RMISE over 200 Monte Carlo Replicates for the Mexican Hat Basis.}  
\resizebox{\textwidth}{!}{\begin{tabular}{ll ccc ccc ccc ccc}  
\toprule \toprule\noalign{\smallskip}

  \\ [-2ex] 
 
& & \multicolumn{6}{c}{$M=2$} & \multicolumn{6}{c}{$M = 4$} \\ 
\cmidrule(l{0.5pt}r{9pt}){3-8}\cmidrule(l{0.5pt}r{9pt}){9-14} 
 & & \multicolumn{3}{c}{Parameters} & \multicolumn{3}{c}{RMISE} & \multicolumn{3}{c}{Parameters} & \multicolumn{3}{c}{RMISE} \\ 
\cmidrule(l{0.5pt}r{9pt}){3-5}\cmidrule(l{0.5pt}r{9pt}){6-8}\cmidrule(l{0.5pt}r{9pt}){9-11}\cmidrule(l{0.5pt}r{9pt}){12-14} 
 $N$ & $q/l_S$ & FKRB & SG & ASG & FKRB & SG & ASG & FKRB & SG & ASG & FKRB & SG & ASG \\

 	\hline \\ [-1ex] 
 
\multicolumn{14}{l}{\textbf{Dimension} $\bm{D=2}$} \\[0.25ex] 
                                                                                                                
  1000 &  3/2 &   9 &  5 & 34.6 & 0.2068 & 0.0724 & 0.0516 &   9 &  5 & 39.6 & 0.1956 & 0.0882 & 0.0577  \\ [0ex] 
  1000 &  7/3 &  49 & 17 & 41.2 & 0.0994 & 0.0484 & 0.0537 &  49 & 17 & 51.0 & 0.1021 & 0.0476 & 0.0595  \\ [0ex] 
  1000 & 15/4 & 225 & 49 & 71.1 & 0.0912 & 0.0482 & 0.0571 & 225 & 49 & 75.1 & 0.0950 & 0.0530 & 0.0625  \\ [0ex] 
 10000 &  3/2 &   9 &  5 & 43.1 & 0.2039 & 0.0706 & 0.0300 &   9 &  5 & 50.5 & 0.1934 & 0.0864 & 0.0435  \\ [0ex] 
 10000 &  7/3 &  49 & 17 & 55.8 & 0.0843 & 0.0424 & 0.0311 &  49 & 17 & 65.4 & 0.0854 & 0.0439 & 0.0422  \\ [0ex] 
 10000 & 15/4 & 225 & 49 & 77.4 & 0.0580 & 0.0326 & 0.0300 & 225 & 49 & 87.0 & 0.0646 & 0.0492 & 0.0397  \\ [0ex]

  \\ [-1ex] 
 
\multicolumn{14}{l}{\textbf{Dimension} $\bm{D=4}$} \\[0.25ex] 
                                                                                                                      
  1000 &  3/2 &   81 &   9 & 153.1 & 0.2254 & 0.0976 & 0.0578 &   81 &   9 & 153.5 & 0.2328 & 0.1199 & 0.0609  \\ [0ex] 
  1000 &  7/3 & 2401 &  49 & 221.5 & 0.1273 & 0.0629 & 0.0589 & 2401 &  49 & 236.4 & 0.1241 & 0.0860 & 0.0613  \\ [0ex] 
  1000 & 15/4 &    $-$ & 209 & 334.1 & $-$ & 0.0495 & 0.0574 &    $-$ & 209 & 338.1 & $-$ & 0.0684 & 0.0576  \\ [0ex]       
 10000 &  3/2 &   81 &   9 & 147.8 & 0.2226 & 0.0972 & 0.0392 &   81 &   9 & 153.8 & 0.2316 & 0.1196 & 0.0488  \\ [0ex] 
 10000 &  7/3 & 2401 &  49 & 211.2 & 0.0787 & 0.0623 & 0.0402 & 2401 &  49 & 228.2 & 0.0915 & 0.0857 & 0.0502  \\ [0ex] 
 10000 & 15/4 &    $-$ & 209 & 339.9 & $-$ & 0.0485 & 0.0364 &    $-$ & 209 & 380.5 & $-$ & 0.0677 & 0.0472  \\ [0ex]

  \\ [-1ex] 
 
\multicolumn{14}{l}{\textbf{Dimension} $\bm{D=6}$} \\[0.25ex] 
                                                                                                                      
  1000 &  3/2 & 729 &  13 &  307.4 & 0.2156 & 0.0850 & 0.0554 & 729 &  13 &  303.2 & 0.2440 & 0.1098 & 0.0674  \\ [0ex] 
  1000 &  7/3 &   $-$ &  97 &  435.6 & $-$ & 0.0644 & 0.0540 &   $-$ &  97 &  451.9 & $-$ & 0.0911 & 0.0605  \\ [0ex]       
  1000 & 15/4 &   $-$ & 545 &  966.7 & $-$ & 0.0609 & 0.0610 &   $-$ & 545 & 1029.4 & $-$ & 0.0866 & 0.0643  \\ [0ex]       
 10000 &  3/2 & 729 &  13 &  292.1 & 0.2138 & 0.0846 & 0.0544 & 729 &  13 &  276.7 & 0.2441 & 0.1095 & 0.0716  \\ [0ex] 
 10000 &  7/3 &   $-$ &  97 &  215.0 & $-$ & 0.0641 & 0.0606 &   $-$ &  97 &  221.0 & $-$ & 0.0909 & 0.0859  \\ [0ex]       
 10000 & 15/4 &   $-$ & 545 & 1052.2 & $-$ & 0.0603 & 0.0680 &   $-$ & 545 & 1098.9 & $-$ & 0.0861 & 0.0574  \\ [0ex]

 	\bottomrule\bottomrule 
 
\end{tabular}} 
\captionsetup{justification=justified,singlelinecheck=true, width=1\textwidth, font={footnotesize,stretch=0.8}, skip=2pt} 
\caption*{\textit{Note:} The table reports the total number of parameter and the RMISE for the FKRB estimator,  the sparse grid estimator (SG), and the adaptive sparse grid estimator (ASG). The adaptive sparse grid estimator performs five refinement steps, whereby the final number of refinements is determined based on the lowest out-of-sample mean squared error calculated with five-fold cross-validation. The grid point to be refined in every refinement step is selected according to its contribution to the local squared error.} 
\end{table} 

\newpage

\begin{figure}[H]
\centering
\captionsetup{skip=0pt, justification=centering, font=normalsize, skip=6pt}
\caption{Probabilities of Further Investments in the Next Six Periods Observed in the Data and Predicted with the Estimators.}
\includegraphics[width=1\textwidth]{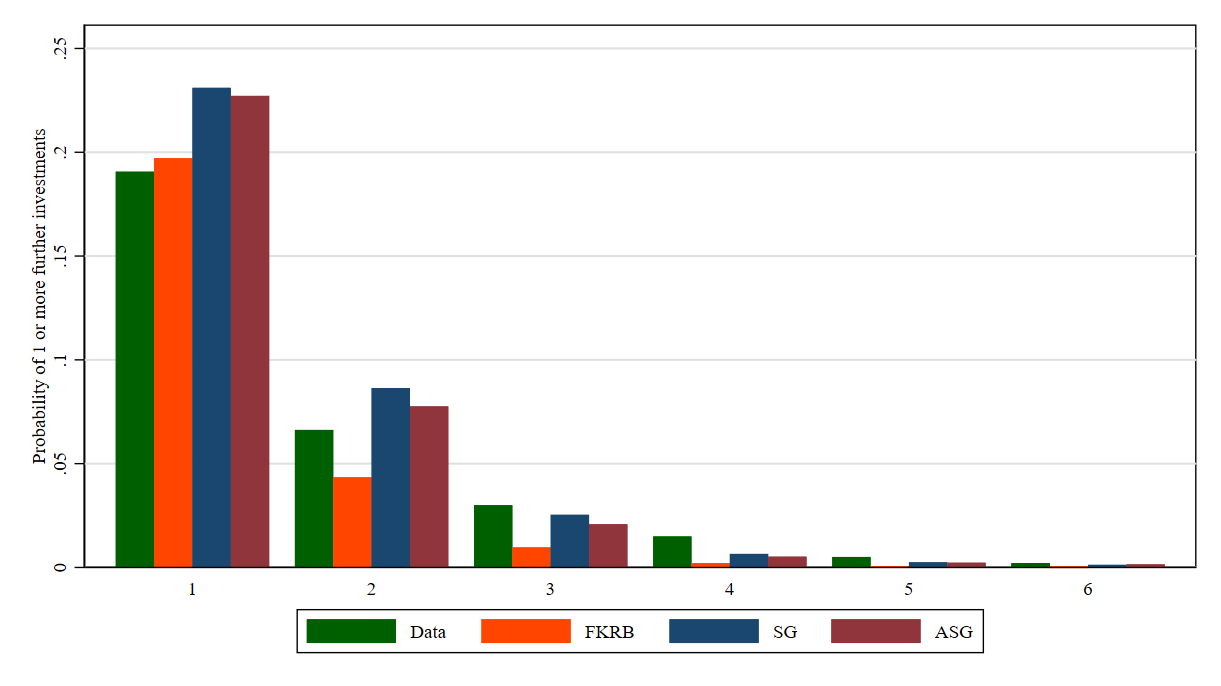}
\captionsetup{justification=justified,singlelinecheck=true, width=0.98\textwidth, font={footnotesize,stretch=0.8}}
\caption{The figure shows the probability of further investments of plants in the six periods after an initial investment observed in data, and predicted by the estimator of \citet{fox2011}, the sparse grid estimator of level $l=4$, and the spatially adaptive sparse grid estimator. The spatially adaptive sparse grid estimator uses ten refinement steps whereby the finale number of refinements is determined using $5$-fold cross-validation and is based on the lowest out-of-sample mean squared error.}
\label{figHB:investments_Application}
\end{figure}

\begin{figure}[H]
\centering
\captionsetup{skip=0pt, justification=centering, font=normalsize, skip=6pt}
\caption{Out-of-sample Mean Squared Error of the Spatially Adaptive Refinement.}
\includegraphics[width=1\textwidth]{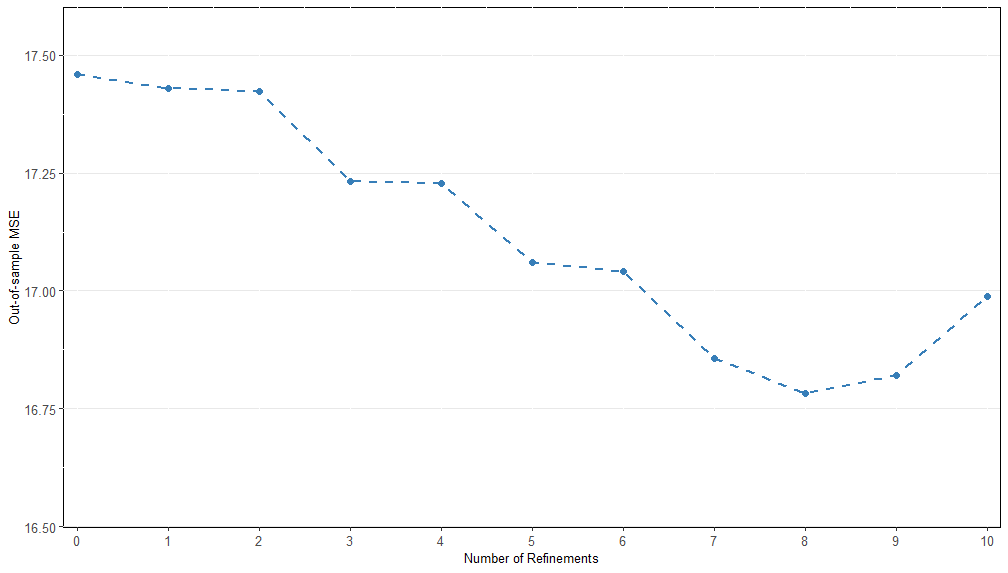}
\captionsetup{justification=justified,singlelinecheck=true, width=0.98\textwidth, font={footnotesize,stretch=0.8}}
\caption{The figure shows the out-of-sample mean squared error (MSE) of the spatially adaptive refinement of the $5$-dimensional sparse grid of level $l_S=4$ calculated via $5$-fold cross-validation. In every refinement step, the grid point with the largest contribution to the squared estimated local error is selected for refinement.}
\label{figHB:application_first_stage_adaptive}
\end{figure}

\newgeometry{left=1.5cm,right=1.5cm,top=2cm,bottom=1.5cm}
\afterpage{

\begin{landscape}

\begin{figure}[H]
\centering
\captionsetup{skip=0pt, justification=centering, font=normalsize, skip=6pt}
\caption{Estimated Histograms of the Five Utility Parameters.}
\includegraphics[width=1.45\textwidth, keepaspectratio=true]{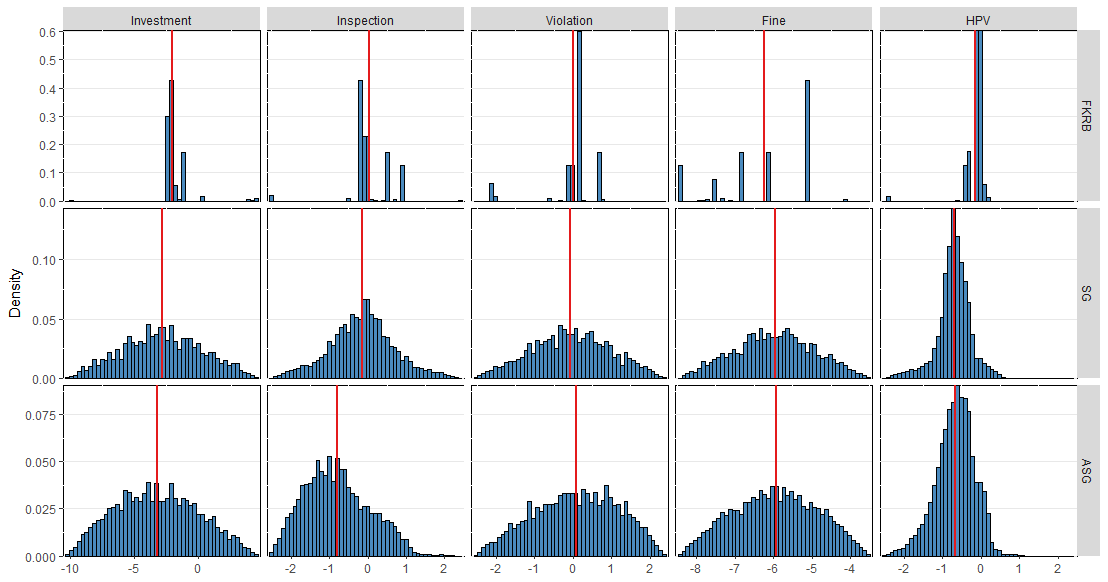}  
\captionsetup{justification=justified,singlelinecheck=true, width=1.4\textwidth, font={footnotesize,stretch=0.8}} 
\caption{The figure shows the histograms for the five utility parameter estimated with the estimator of \citet{fox2011} (FKRB), wit the sparse grid estimator (SG), and the spatially adaptive sparse grid estimator (ASG). The red lines show the means of the estimated distribution in every of the five dimensions. The spatially adaptive sparse grid estimator uses ten refinement steps whereby the finale number of refinements is determined using $5$-fold cross-validation and based on the lowest out-of-sample mean squared error.} \label{figHB:hist_application}
\end{figure}

\end{landscape}	

}

\restoregeometry
\newpage

\end{appendix}

\end{document}